\documentclass[aps,twocolumn,pra,showpacs,superscriptaddress,amsmath,amssymb,showkeys,10pt]{revtex4-1}
\usepackage{graphicx}
\usepackage{epstopdf}
\usepackage{braket}
\usepackage{hyperref}
\usepackage{wrapfig}
\allowdisplaybreaks

\begin{document}

\title{Visualizing the breakdown of quantum multimodality in coherently driven light-matter interaction}

\author{Th. K. Mavrogordatos}
\email[Email addresses: ]{themis.mavrogordatos@fysik.su.se;\\ themis.mavrogordatos@icfo.eu}
\affiliation{Department of Physics, Stockholm University, SE-106 91 Stockholm, Sweden}
\affiliation{ICFO -- Institut de Ci\`{e}ncies Fot\`{o}niques, The Barcelona Institute of Science and Technology, 08860 Castelldefels (Barcelona), Spain}

\date{\today}

\begin{abstract}
We show that the saturation of a multiphoton transition is accompanied by a gradual collapse of quantum multimodality in the strong-coupling limit of the weakly driven Jaynes-Cummings (JC) model. By means of a perturbative method, we illustrate the prominent role of quantum fluctuations by focusing on the two and three-photon resonance operation in a regime where the steady-state average photon number is below or marginally above unity. We also reveal two coexistent quantum beats in the intensity correlation function of the forwards scattered photons. These beats, originating from the states mediating the cascaded decay, arise as a direct consequence of the distinct JC spectrum. Their interference coordinates with the alternation of positive and negative values of the Wigner function around the phase-space origin in the transient conditioned on a photodetection.    
\end{abstract}

\pacs{42.50.Hz, 42.50.Ar, 42.50.-p}
\keywords{multiphoton resonances, dressing of dressed states, quantum-classical correspondence, Wigner distribution, second-order correlation, quantum trajectories, cavity and circuit QED, Jaynes-Cummings model}

\maketitle

\section{Introduction}
The response of fluctuating nonlinear oscillators with no analog in thermal equilibrium systems has guided the development of several branches in modern physics, from the manifestation of collective dynamical phenomena in coupled mechanical resonators~\cite{NEMS2012, NanowiresNE} to the investigation of dissipative quantum phase transitions in zero dimensions in circuit quantum electrodynamics (circuit QED)~\cite{Fink2017}. A pivotal direction among these ramifications is enabled by current experimental advancements in quantum optics focusing on operating regimes where the treatment of fluctuations can no longer rely on concepts adapted to the small-noise limit~\cite{CarmichaelQO2}. A characteristic example in this direction is provided by the occurrence photon blockade as a dynamical phenomenon in driven systems with an anharmonic spectrum~\cite{Imamoglu1997, Hamsen2017, Miranowicz2019}; the absorption of $n$ photons establishes an effective two-level structure which blocks the absorption of an additional incoming $n+1$ photon. This effect is naturally linked to photon antibunching and sub-Poissonian statistics~\cite{Carmichael1976, Kimble1977, Dagenais1978, Short1983, Birnbaum2005, Lang2011}, arising as distinct quantum features.

Not only is the photon blockade dynamical, but also its breakdown, relying in a fundamental way on the open character of the system, as does the occurrence of single-atom bistability -- the mechanism by means of which photon blockade breaks down -- first reported in 1988~\cite{SingleAtomBist} and subsequently demonstrated in a beautiful experiment by Kerckhoff and coworkers~\cite{Mabuchi2011} in the presence of detuning, using a homodyne measurement scheme. Photon blockade breakdown, alongside its link to spontaneous symmetry breaking in a second-order quantum phase transition with critical scaling on resonance\cite{Alsing1991, Mabuchi2009, Curtis2021} -- anticipated by the collapse of the quasi-energy spectrum at threshold~\cite{Alsing1992} -- explicitly uncovers the role of one, two, three or more quanta in producing energy level shifts and multiphoton transitions that are missed from the semiclassical nonlinear response in the presence of detuning. It is this very regime which is subject to a strong-coupling ``thermodynamic limit'' where quantum fluctuations persist and multiphoton resonances of higher order appear, turning increasingly sharper as the associated system-size parameter -- scaling with the square of the ratio between the light-matter interaction strength and the cavity decay rate -- tends to infinity~\cite{PhotonBlockade2015}.   
\begin{figure*}
 \centering
 \includegraphics[width=0.92\textwidth]{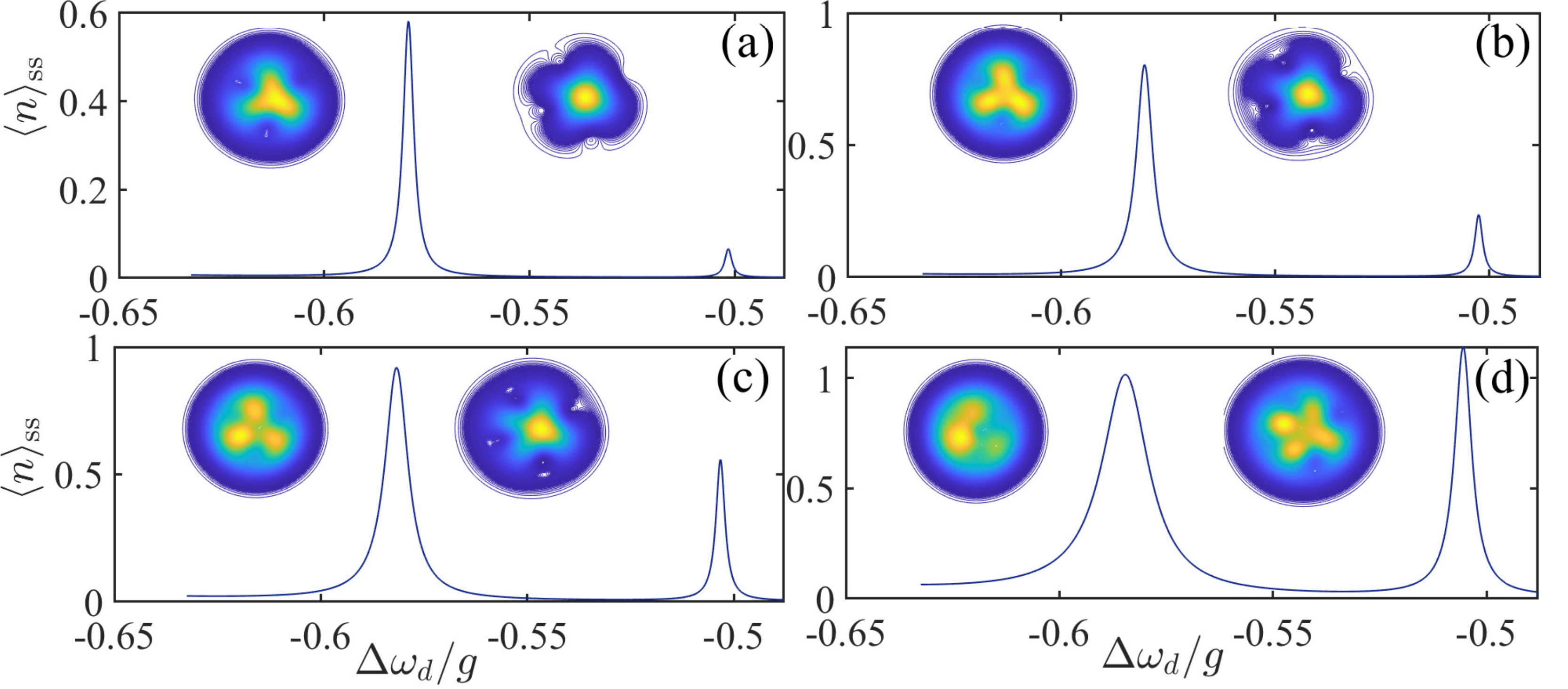}
 \caption{{\it From tri to quad-modality across adjacent resonances}. Average photon number in the steady state, $\braket{n}_{\rm ss}$, obtained from the solution of the master equation for the driven and damped JC model, as a function of the dimensionless drive detuning $\Delta\omega_d/g$ for $\varepsilon_d/g=0.05, 0.06, 0.07$ and $0.09$ in frames {\bf (a)}--{\bf (d)}, respectively. The left (right) peak in each frame corresponds to the three (four) photon resonance, while the left (right) inset depicts a schematic contour plot of the Wigner distribution for the steady-state intracavity field at the peak location of the three (four) photon resonance. We take $g/\gamma=500$ and $\gamma/(2\kappa)=1$.}
 \label{fig:introfig}
\end{figure*}

The two-photon resonance for a single atom trapped in a high-finesse optical cavity was studied in 2008 at the focus of an experiment that ``enters a new regime, with nonlinear quantum optics at the level of individual atomic and photonic quanta''~\cite{Schuster2008}. In this report, the authors commented that ``according to bistability theory, we are operating on the lower branch, where the corresponding nonlinear response is small...Indeed, the reported nonlinearity occurs with an occupation probability of the atomic excited state of at most $0.07$. This is what makes it radically different from and dominant over the standard saturation nonlinearity for a two-state atom''. Shortly after this demonstration, Bishop and collaborators~\cite{Bishop2009} provided direct evidence on the extended Jaynes-Cummings (JC) energy spectrum in one of the most characteristic experiments of circuit QED [see also the preceding experimental evidence of the $\sqrt{n}$ JC nonlinearity~\cite{Fink2008}], and built upon the early concepts of photon blockade~\cite{Tian1992, CarmichaelQO2} to interpret vacuum Rabi splitting, five years after the vacuum Rabi frequency had been shown to exceed the qubit and cavity damping rates~\cite{Blais2004}. In the quest of light sources producing bound states of multiple photons~\cite{Chang2014}, the nonclassical properties evinced by third and fourth-order correlation functions were used to assess the multiphoton states of light emanating from a photonic crystal nanocavity strongly coupled to a quantum dot~\cite{Faraon2008, Majumdar2012, Rundquist2014}. As well as appealing to higher-order correlations, the first experimental observation of the two-photon blockade for a single $ ^{87}$Rb atom interacting with a resonant mode of a high-finesse Fabry-P\'{e}rot cavity~\cite{Hamsen2017} came with the violation of the Cauchy-Schwarz inequality as the intensity correlation function falls below its zero-delay value. 

From the above background, it becomes clear that some key pieces are missing to complete the puzzle of what lies behind a multiphoton resonance established in a quantum nonlinear oscillator with light-matter coupling. Prompted by the explanation in the above paragraph provided by Schuster and collaborators in~\cite{Schuster2008} as well as by the recent demonstration of bimodality in the JC two-photon resonance~\cite{Wigner2PB} and some of the multimodal profiles presented in~\cite{Mavrogordatos2019} when asymptotically assessing the strong-coupling limit of light-matter interaction, we may ask again what is the place of multistability in what appears to be the saturation of an effective two-level transition? How are we to understand the nonclassical nature of such an output photon stream from the visualization of fluctuations within the quantum-classical correspondence in the language of {\it quasi}probability distributions~\cite{CarmichaelQO1, CarmichaelQO2}? To set the scene, Fig.~\ref{fig:introfig} depicts the collapse of tri-modality in favor of a developing quad-modality for increasing drive strength, exposing a background of nonlinearity which deviates substantially from the predictions of mean-field theory.

In this article, we delve into the phase-space representation of the electromagnetic field in the open driven JC model deep into the strong-coupling limit. We will find that the saturation of a multiphoton resonance comes along with the collapse of multimodality in a region where quantum fluctuations leave an indelible mark. This mark is also imprinted on the second-order coherence of the emitted radiation. After formulating the secular approximation in the dressed-state basis in Sec.~\ref{sec:MEdressedstate}, we derive the rate equations for the matrix elements of the system density operator in Sec.~\ref{sec:rateeqs}. These results are used for the determination of the {\it quasi}probability distribution of the intracavity field in Sec.~\ref{sec:WignSST}, and subsequently for the calculation of the intensity correlation function for the forwards scattered photons in Sec.~\ref{sec:2ndordcoh}. Our main findings are discussed in light of their experimental verification in current setups of multiphoton quantum nonlinear optics, before being briefly summarized in the Conclusion.  

\section{Master equation in the dressed-state basis}
\label{sec:MEdressedstate}

Within the standard description, the system density matrix $\rho$ obeys the Lindblad master equation (ME)
\begin{align}\label{eq:ME1}
 \frac{d\rho}{dt}&=\mathcal{L}\rho \equiv -i[\omega_0(\sigma_{+}\sigma_{-} + a^{\dagger}a)+g(a\sigma_{+}+a^{\dagger}\sigma_{-}),\rho]\notag \\
 &-i[ (\varepsilon_d^{*} a e^{i\omega_d t} + \varepsilon_d a^{\dagger}e^{-i\omega_d t}),\rho]\notag \\
 &+\kappa (2 a \rho a^{\dagger} -a^{\dagger}a \rho - \rho a^{\dagger}a)\notag \\
 &+\frac{\gamma}{2}(2\sigma_{-}\rho \sigma_{+} - \sigma_{+}\sigma_{-}\rho - \rho \sigma_{+}\sigma_{-}),
\end{align}
where $a$ and $a^{\dagger}$ are the annihilation and creation operators for the cavity photons, $\sigma_{+}$ and $\sigma_{-}$ are the raising and lowering operators for the two-level atom, $g$ is the dipole coupling strength, $2\kappa$ is the rate at which photons are lost from the cavity, and $\gamma$ is the spontaneous emission rate for the atom to modes other than the privileged cavity mode. The cavity is coherently driven with a field of amplitude $\varepsilon_d$ (assumed to be real, without loss of generality) and frequency $\omega_d$ defining the detuning $\Delta \omega_d \equiv \omega_d-\omega_0$. Hereinafter, we operate in the strong-coupling limit with $g/\kappa=10^3$ in keeping with~\cite{Shamailov2010, Bishop2009}, while we select $\gamma/(2\kappa)=1$ as an impedance matching condition, which considerably simplifies the analysis (see~\cite{CarmichaelQO2, Shamailov2010}). 

\begin{figure}
 \centering
 \includegraphics[width=0.35\textwidth]{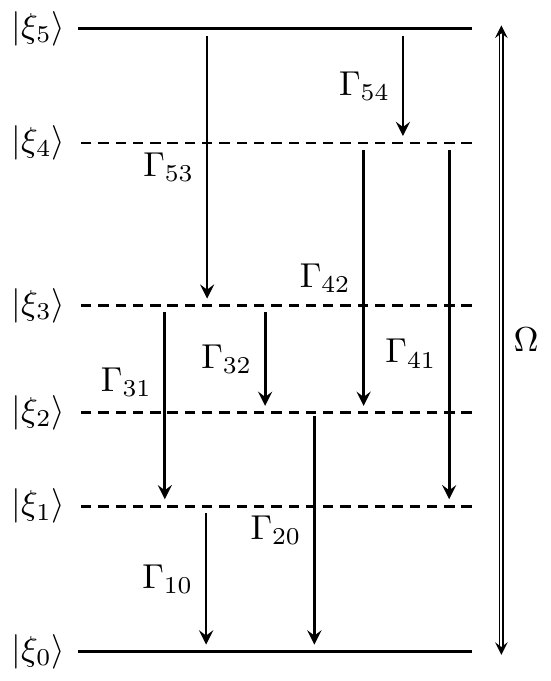}
 \caption{Schematic representation of the effective model displaying six shifted dressed-state levels. The three-photon Rabi frequency is denoted by $\Omega$ while the displayed levels are shifted energy levels (see the Appendix). The intermediate levels $\ket{\xi_1}, \ket{\xi_2}, \ket{\xi_3}, \ket{\xi_4}$, marked by dashed lines, are occupied via the cascaded decay and are responsible for the formation of the two quantum beats of partial visibility.}
 \label{fig:6levels}
\end{figure}

In the strong-coupling limit ($g \gg \kappa, \gamma/2$) the problem reduces to a minimal model accounting for the multiphoton transition which is resonantly excited in the weak excitation regime. Such a perturbative treatment, formally developed under the guidance of~\cite{GottfriedQM}, was derived in~\cite{Shamailov2010} on the basis of adiabatic elimination of intermediate states to assess the first and second-order coherence of the forwards scattered light for the two-photon resonance in one-atom cavity QED, essentially employs an expansion in powers of $\varepsilon_d/g \ll 1 $. Our interest here is primarily with the three-photon resonance for which we employ the model depicted in Fig.~\ref{fig:6levels}. The first six JC dressed states between which transitions are constrained to occur, starting from the ground state, are denoted by
\begin{subequations}\label{eq:6levels}
\begin{align}
 &\ket{\xi_0}\equiv\ket{0,-}, \\
 &\ket{\xi_1}\equiv \frac{1}{\sqrt{2}} (\ket{1,-} - \ket{0,+}), \\
 & \ket{\xi_2}\equiv \frac{1}{\sqrt{2}} (\ket{1,-} + \ket{0,+}), \\
 &  \ket{\xi_3}\equiv \frac{1}{\sqrt{2}} (\ket{2,-} - \ket{1,+}), \\
  &  \ket{\xi_4}\equiv \frac{1}{\sqrt{2}} (\ket{2,-} + \ket{1,+}), \\
   &  \ket{\xi_5}\equiv \frac{1}{\sqrt{2}} (\ket{3,-} - \ket{2,+}),
\end{align}
\end{subequations}
where $\ket{n, \pm}\equiv \ket{n}\otimes\ket{\pm}$, $\ket{n}$ is the Fock state of the cavity field, while $\ket{+}, \ket{-}$ are the upper and lower states of the two-level atom, respectively. We now proceed to make the secular approximation~\cite{Cohen_Tannoudji_1977, CarmichaelQO2} leading to the following effective ME in the laboratory frame [see~\cite{Lledo2021}, Eq. (18) of~\cite{Shamailov2010} and Eq. (16) of~\cite{Beaudoin2011}]:
\begin{equation}\label{eq:ME2}
\begin{aligned}
  &\frac{d\rho}{dt}= -i[\tilde{H}_{\rm eff},\rho]+\Gamma_{54} \mathcal{D}[|\xi_4\rangle \langle \xi_5|](\rho)\\
  & + \Gamma_{53} \mathcal{D}[|\xi_3\rangle \langle \xi_5|](\rho)+ \Gamma_{42} \mathcal{D}[|\xi_2\rangle \langle \xi_4|](\rho) + \Gamma_{41} \mathcal{D}[|\xi_1\rangle \langle \xi_4|](\rho)\\
  & + \Gamma_{32} \mathcal{D}[|\xi_2\rangle \langle \xi_3|](\rho) + \Gamma_{31} \mathcal{D}[|\xi_1\rangle \langle \xi_3|](\rho)+ \Gamma_{20} \mathcal{D}[|\xi_0\rangle \langle \xi_2|](\rho)\\
  &+ \Gamma_{10} \mathcal{D}[|\xi_0\rangle \langle \xi_1|](\rho),
  \end{aligned}
\end{equation}
where the effective Hamiltonian governing the coherent dynamical evolution is
\begin{equation}\label{eq:EffHam}
 \tilde{H}_{\rm eff}\equiv\sum_{k=0}^{5} \tilde{E}_{k} |\xi_k\rangle \langle \xi_k| + \hbar \Omega (e^{3i\omega_d t} |\xi_0\rangle \langle \xi_5| + e^{-3i\omega_d t} |\xi_5\rangle \langle \xi_0|),
\end{equation}
with $ \tilde{E}_k=E_k + \hbar\delta_k(\varepsilon_d)$ ($k=0, 1,\ldots, 5$) the JC eigen-energies dressed by the interaction with the drive. In the effective ME of Eq.~\eqref{eq:ME2}, we define as usual $\mathcal{D}[X](\rho)\equiv X\rho X^{\dagger}-(1/2)\{X^{\dagger}X, \rho\}$. The transition rates between the dressed states, within the secular approximation, read
 \begin{subequations}
\begin{align}
 \Gamma_{10}&=\Gamma_{20}=\frac{1}{2}\gamma + \kappa, \\
 \Gamma_{31}&=\Gamma_{42}=\frac{1}{4}\gamma + \frac{1}{2} \kappa (\sqrt{2}+1)^2, \\
 \Gamma_{32}&=\Gamma_{41}=\frac{1}{4}\gamma + \frac{1}{2} \kappa (\sqrt{2}-1)^2, \\
  \Gamma_{53}&=\frac{1}{4}\gamma + \frac{1}{2} \kappa (\sqrt{3}+\sqrt{2})^2, \\
 \Gamma_{54}&=\frac{1}{4}\gamma + \frac{1}{2} \kappa (\sqrt{3}-\sqrt{2})^2.
\end{align}
\end{subequations}
For the operating conditions we will be working with, second and third powers of the ratio $\varepsilon_d/g$ may be comparable in size to $\kappa/g$ or $\gamma/g$; this instance justifies keeping the first-order terms in the dissipation rates that are produced by super-operators of the Lindblad form. Explicit expressions for the transition rates between dressed states, $\Gamma_{ij}$, and the energy shifts $\delta_k$ are given in Sec.~\ref{sec:rateeqs} and the Appendix. The three-photon Rabi frequency derives from the sum of four pathways connecting the states $\ket{\xi_5}$ and $\ket{\xi_0}$ with non-zero matrix elements of the perturbation $\varepsilon_d(a^{\dagger}+a)$, yielding $\Omega  \approx 11.7\, \varepsilon_d^3/g^2$. The three-photon resonance must be excited with a drive detuning given by $\Delta\omega_d=-g/\sqrt{3} + (1/3)(\delta_5-\delta_0)$, which is now an explicit function of the drive strength. To truncate the ladder of the JC dressed states and form our minimal model, we also note that the drive frequency for three-photon absorption, $\omega_0-g/\sqrt{3}$, is far from resonance for the occurrence of four-photon absorption, a condition which requires $(4-2\sqrt{3})g \gg \sqrt{3}(\kappa,\gamma/2)$.

\section{Rate equations in the secular approximation}
\label{sec:rateeqs}

In this section, we detail the six-level effective model upon which our perturbative treatment relies. The transition rates involved in the dynamics are expanded in powers of $\varepsilon_d/g \ll 1$, while we always operate in the strong-coupling limit with $g/\kappa=10^3$, within the secular approximation. To simplify the calculations we consider the special case $\gamma=2\kappa$, for which the frequency of the vacuum Rabi oscillation is independent of the decoherence rates~\cite{CarmichaelQO2}. We will use our results to the calculation of the intensity correlation function for the forwards-scattered photons, and the steady-state and transient Wigner distributions of the intracavity field. 

We start with recasting the system operators in the dressed-state formalism. The annihilation operator in the dressed-state basis, limited to the minimal set of orthogonal dressed states $\ket{\xi_0} -\ket{\xi_5}$, is 
\begin{equation}
\begin{aligned}
 a& \approx \frac{1}{\sqrt{2}} |\xi_0 \rangle \langle \xi_1| +\frac{1}{\sqrt{2}} |\xi_0 \rangle \langle \xi_2| + \frac{\sqrt{2}+1}{2} |\xi_1 \rangle \langle \xi_3|\\
 &+ \frac{\sqrt{2}-1}{2} |\xi_2 \rangle \langle \xi_3| + \frac{\sqrt{2}+1}{2} |\xi_2 \rangle \langle \xi_4| + \frac{\sqrt{2}-1}{2} |\xi_1 \rangle \langle \xi_4|\\
 &+ \frac{\sqrt{3}+\sqrt{2}}{2} |\xi_3 \rangle \langle \xi_5| + \frac{\sqrt{3}-\sqrt{2}}{2} |\xi_4 \rangle \langle \xi_5|.
 \end{aligned}
\end{equation}
Taking the conjugate, we find that the photon-number operator assumes the form
\begin{equation}\label{eq:photonopdr}
\begin{aligned}
 & a^{\dagger}a \approx (1/2)\left(|\xi_1\rangle \langle \xi_1|+|\xi_2\rangle \langle \xi_2|+|\xi_2\rangle \langle \xi_1| + |\xi_1\rangle \langle \xi_2| \right) \\
 &+ (3/2) (|\xi_3\rangle \langle \xi_3| +  |\xi_4\rangle \langle \xi_4|)+(1/2)\left(|\xi_3\rangle \langle \xi_4| + |\xi_4\rangle \langle \xi_3|\right)\\
 &+ (5/2) |\xi_5\rangle \langle \xi_5|.
 \end{aligned}
\end{equation}
To assess the second-order coherence of the transmitted light we require the intensity correlation function~\cite{CarmichaelQO2, OpenSystems2013} (the steady state is denoted by the subscript ss),
\begin{equation}
\begin{aligned}
g^{(2)}_{\rightarrow}(\tau)&=\frac{\braket{a^{\dagger}(0)a^{\dagger}(\tau)a(\tau)a(0)}_{\rm ss}}{\braket{a^{\dagger}a}_{\rm ss}^2} \\
&=\frac{{\rm tr}\{a^{\dagger}(0)a(0) e^{\mathcal{L}\tau}[a(0)\rho_{\rm ss}a^{\dagger}(0)]\}}{\braket{a^{\dagger}a}_{\rm ss}^2}\\
&=\frac{\braket{(a^{\dagger}a)(\tau)}_{\rho(0)=\rho_{\rm cond}}}{\braket{a^{\dagger}a}_{\rm ss}},
\end{aligned}
\end{equation}  
with reference to the ME~\eqref{eq:ME1}. A photon emitted from the cavity after the steady state is attained creates the superposition quantum states
\begin{equation}\label{eq:superst}
\begin{aligned}
 \ket{\xi_3} &\to \ket{\psi_{\rm super,\,1}}=\sqrt{\frac{2}{3}}\left(\frac{\sqrt{2}+1}{2}\ket{\xi_1} +  \frac{\sqrt{2}-1}{2}\ket{\xi_2}\right), \\
 \ket{\xi_5} &\to \ket{\psi_{\rm super,\,2}}=\sqrt{\frac{2}{5}}\left(\frac{\sqrt{3}+\sqrt{2}}{2}\ket{\xi_3} +  \frac{\sqrt{3}-\sqrt{2}}{2}\ket{\xi_4}\right),
 \end{aligned}
\end{equation}
and also maps the population of the first-excited couplet to the ground state. This is how the intermediate states $\ket{\xi_1}$ -- $\ket{\xi_4}$ enter the dynamics in the form of two quantum beats -- in other words, two coherent superpositions -- originating from two JC couplets. Therefore, as a result of conditioning on a photon emission, the following density operator is prepared,
\begin{widetext}
\begin{equation}\label{eq:rhocondit}
\begin{aligned}
 \rho_{\rm cond}&=\frac{1}{\braket{a^{\dagger}a}_{\rm ss}}\left[\frac{1}{2}(p_1+p_2)|\xi_0\rangle \langle \xi_0| + \frac{3}{2} (p_3 + p_4) |\psi_{\rm super,\,1}\rangle \langle \psi_{\rm super,\,1}| + \frac{5}{2} p_5 |\psi_{\rm super,\,2}\rangle \langle \psi_{\rm super,\,2}|\right]\\
 &=\frac{6}{25} |\xi_0\rangle \langle \xi_0| + \frac{9}{25} |\psi_{\rm super,\,1}\rangle \langle \psi_{\rm super,\,1}| + \frac{10}{25} |\psi_{\rm super,\,2}\rangle \langle \psi_{\rm super,\,2}|,
 \end{aligned}
\end{equation}
\end{widetext}
where the steady-state occupation probabilities of the six dressed energy levels denoted by $p_i \equiv (\rho_{ii})_{\rm ss}$, $i=0, 1,\ldots, 5$. Taking the matrix elements with respect to the dressed-state basis, we find
\begin{equation}\label{eq:g2}
\begin{aligned}
 &g_{\rightarrow}^{(2)}(\tau)= \frac{{\rm tr} \left\{[e^{\mathcal{L}\tau}\rho_{\rm cond}] a^{\dagger}a\right\}}{\braket{a^{\dagger}a}_{\rm ss}}= \frac{\braket{(a^{\dagger}a)(\tau)}}{{\braket{a^{\dagger}a}_{\rm ss}}} \Bigg|_{\rho(0)=\rho_{\rm cond}}.
 \end{aligned}
\end{equation}
In terms of the matrix elements in the dressed-state basis,
\begin{equation*}
\begin{aligned}
 &\braket{(a^{\dagger}a)(\tau)}=\frac{1}{2}[\rho_{11}(\tau)+\rho_{22}(\tau)+\rho_{12}(\tau)+\rho_{21}(\tau)\\
 &+\rho_{34}(\tau)+\rho_{43}(\tau)] + \frac{3}{2}[\rho_{33}(\tau)+\rho_{44}(\tau)] + \frac{5}{2} \rho_{55}(\tau).
\end{aligned}
 \end{equation*}

We will now show that the photon number in the steady state is
\begin{equation}
 \braket{a^{\dagger}a}_{\rm ss}=\frac{1}{2}(p_1+p_2) + \frac{3}{2}(p_3 + p_4) + \frac{5}{2}p_5,
\end{equation}
where we have used Eq.~\eqref{eq:photonopdr}. To this end, we determine the transition rates featuring in Fig.~\ref{fig:6levels}. Invoking the principle of detailed balance (see also Sec. 3.2 of~\cite{Shamailov2010}), we obtain
\begin{subequations}
\begin{align}
 \Gamma_{10} p_1&=\Gamma_{31}p_3 + \Gamma_{41}p_4, \\
 \Gamma_{20} p_2&=\Gamma_{32} p_3 + \Gamma_{42} p_4, \\
 \Gamma_{53} p_5&=(\Gamma_{31} + \Gamma_{32})p_3, \\
 \Gamma_{54} p_5&=(\Gamma_{42} + \Gamma_{41}) p_4,
\end{align}
\end{subequations}
For the special case $\gamma=2\kappa$, the photon number in the steady state in terms of $p_5$ is
\begin{equation}
  \braket{a^{\dagger}a}_{\rm ss}=\frac{1}{2}\cdot 3p_5 + \frac{3}{2} \cdot \frac{3}{2} p_5+ \frac{5}{2}p_5=\frac{25}{4}p_5.
\end{equation}
The equations of motion for the matrix elements involved read (the dot means the derivative with respect to $\tau$)
\begin{subequations}\label{eq:syseq1}
\begin{align}
 \dot{\rho}_{00}&=\Gamma_{10} \rho_{11}  + \Gamma_{20} \rho_{22}-i\Omega (\rho_{50}e^{3i\omega_d \tau}-\rho_{05}e^{-3i\omega_d \tau}), \\
  \dot{\rho}_{11}&=-\Gamma_{10} \rho_{11} + \Gamma_{31} \rho_{33} + \Gamma_{41} \rho_{44}, \\
   \dot{\rho}_{22}&=-\Gamma_{20} \rho_{22} + \Gamma_{32} \rho_{33} + \Gamma_{42} \rho_{44}, \\
    \dot{\rho}_{33}&=-(\Gamma_{31} + \Gamma_{32}) \rho_{33} + \Gamma_{53} \rho_{55}, \\
    \dot{\rho}_{44}&=-(\Gamma_{42}+\Gamma_{41}) \rho_{44} + \Gamma_{54} \rho_{55}, \\
    \dot{\rho}_{55}&=-(\Gamma_{53}+\Gamma_{54})\rho_{55} -i \Omega (\rho_{05}e^{-3i\omega_d \tau}-\rho_{50}e^{3i\omega_d \tau}), \\
     \dot{\rho}_{12}&= \dot{\rho}_{21}^{*}=-(i/\hbar)(\tilde{E}_1 - \tilde{E}_2) \rho_{12} 
     - \frac{\Gamma_{10}+\Gamma_{20}}{2}\rho_{12}, \label{eq:beat1}\\
     \dot{\rho}_{34}&= \dot{\rho}_{43}^{*}=-(i/\hbar)(\tilde{E}_3 - \tilde{E}_4) \rho_{34}\notag\\
     &- \frac{\Gamma_{31}+\Gamma_{41}+\Gamma_{32}+\Gamma_{42}}{2}\rho_{34}, \label{eq:beat2}\\
     \dot{\rho}_{05}&=\dot{\rho}_{50}^{*}=-(i/\hbar)(\tilde{E}_0 - \tilde{E}_5) \rho_{05}\notag\\
     &- i\Omega(\rho_{55}-\rho_{00})e^{3i\omega_d \tau} - \frac{\Gamma_{53}+\Gamma_{54}}{2}\rho_{05}.
  \end{align}
   \end{subequations}
We now transform to a frame rotating with the drive via an appropriate unitary transformation. The drive frequency must be tuned to the three-photon transition, determined by
\begin{equation}\label{eq:rescond}
 3\omega_d=3\omega_0 - \sqrt{3}g + (\delta_5-\delta_0),
\end{equation}
where the shifts $\delta_{i}$ ($i=0,1,\ldots,5$) account for the dressing of the energy levels by the drive; we will calculate them in the next section. Noting that $\tilde{E}_5 - \tilde{E}_0=\hbar[3\omega_0 - \sqrt{3}g + (\delta_5-\delta_0)]$, the coupled subset of Eqs.~\eqref{eq:syseq1} is then conveniently recast in the form [$D \equiv {\rm Im}(\rho_{05})$]
\begin{subequations}\label{eq:syseq2}
\begin{align}
&\dot{\rho}_{00}=\gamma(\rho_{11}  + \rho_{22})-2\Omega D, \\
 &\dot{\rho}_{11}+ \dot{\rho}_{22}=-\gamma(\rho_{11} + \rho_{22}) + 2\gamma (\rho_{33} + \rho_{44}), \\
  &\dot{\rho}_{33}+ \dot{\rho}_{44}=-2\gamma(\rho_{33} + \rho_{44}) + 3\gamma \rho_{55}, \\
   &\dot{\rho}_{55}=-3\gamma\rho_{55} +2\Omega D, \\
   &\dot{D}=-\Omega(\rho_{55}-\rho_{00}) - \frac{3}{2}\gamma D,
  \end{align}
  \end{subequations}
with initial conditions $\rho_{00}(0)=6/25$, $D(0)=\rho_{55}(0)=0$, $\rho_{11}(0)+ \rho_{22}(0)=9/25$, $\rho_{33}(0)+ \rho_{44}(0)=10/25$. From the steady-state solution of Eqs.~\eqref{eq:syseq2}, we obtain the excitation probability
\begin{equation}
 p_5=\frac{4\Omega^2}{26\Omega^2 + 9 \gamma^2},
\end{equation}
with the three-photon Rabi frequency $\Omega$ given in Eq.~\eqref{eq:Omega3p} of the Appendix. Compare to the corresponding expression for the two-photon resonance~\cite{Shamailov2010, Lledo2021},
\begin{equation}
 p_3=\frac{\Omega^{\prime 2}}{4 \Omega^{\prime 2} +\gamma^2},
\end{equation}
where the $p_3$ is the steady-state occupation probability of the upper level (the state $\ket{\xi_3}$) and $\Omega^{\prime}=2\sqrt{2} \varepsilon_d^2/g$. We then determine the intensity correlation at zero delay as
\begin{equation}\label{eq:g20analytical}
 g^{(2)}_{\rightarrow}(\tau=0)=\frac{\braket{a^{\dagger}a}|_{\rho_{\rm cond}(0)}}{\braket{a^{\dagger}a}_{\rm ss}}=\frac{22}{25}\cdot \frac{4}{25 p_5},
\end{equation}
with $g_{\rightarrow}^{(2)}(\tau=0) \to 0.915$ for $(\Omega/\gamma)^2 \gg 1$. In the opposite limit, for $(\Omega/\gamma)^2 \ll 1$, photon emission is highly bunched, which is also the case for the two-photon resonance~\cite{Shamailov2010}.
\begin{figure*}
\centering
\includegraphics[width=0.81\textwidth]{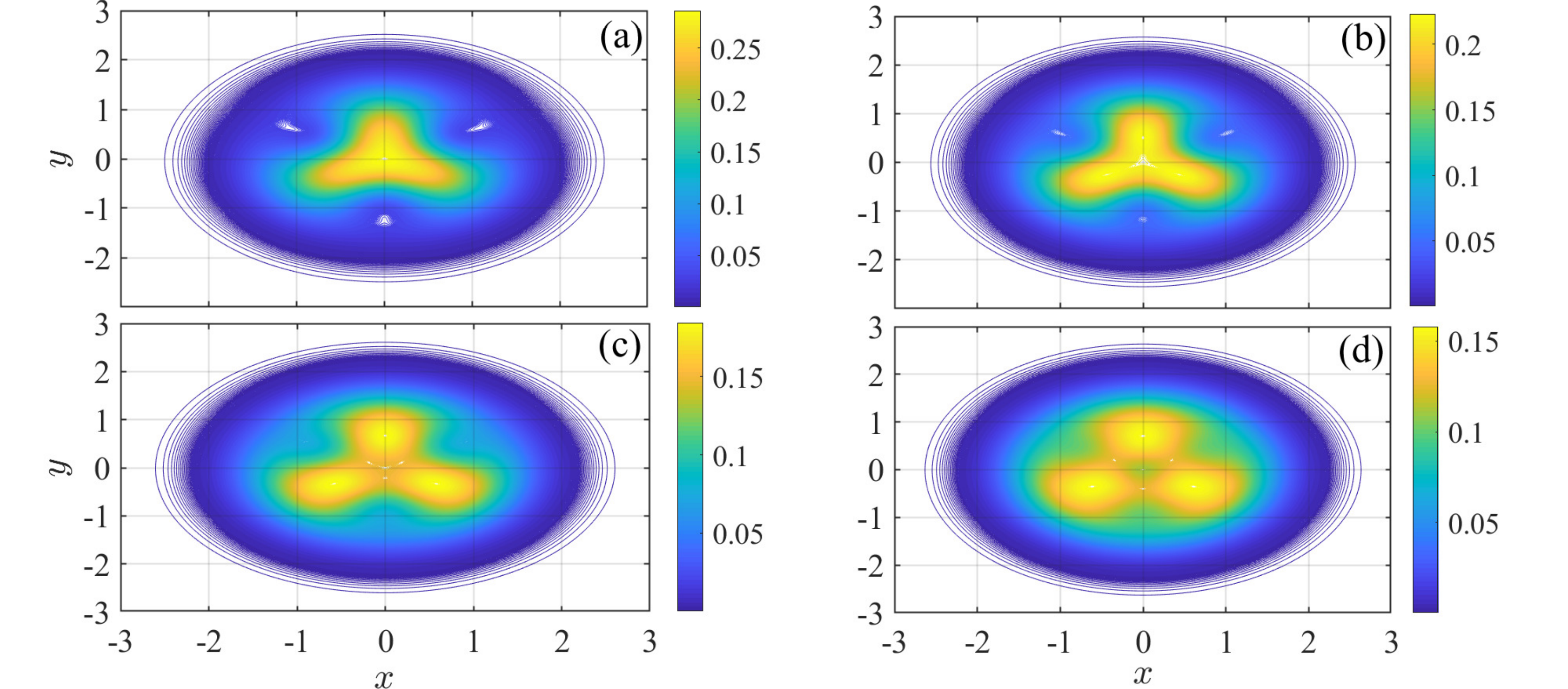}
\caption{{\it Development of steady-state trimodality in the minimal six-level model.} Wigner function (contour plots) of the intracavity field (in the frame rotating with the drive), $W_{\rm ss}(x+iy)$, calculated from Eq.~\eqref{eq:WssSA} for the detuning of Eq.~\eqref{eq:rescond}. The steady-state excitation probability of the state $\ket{\xi_5}$ is $p_5=0.10, 0.12, 0.14$ and $0.15$ in frames {\bf (a)}--{\bf (d)}, respectively, while $g/\gamma=500$ and $\gamma=2\kappa$. All color bars here and henceforth indicate the values of the Wigner function.}
\label{fig:Wssanalytical}
\end{figure*}

We now turn to the system of Eqs.~\eqref{eq:syseq1}. The equations involving diagonal matrix elements which are not coupled to the drive can be readily solved. From Eq.~\eqref{eq:beat1} and the expression for $\rho_{\rm cond}(0)$, it is evident that
\begin{equation}
 \rho_{12}(\tau)=\rho_{12}(0) e^{-\gamma \tau} e^{i \nu_1 \tau}=\frac{3}{50} e^{-\gamma \tau} e^{i \nu_1 \tau},
\end{equation}
while from Eq.~\eqref{eq:beat2} we obtain
\begin{equation}
 \rho_{34}(\tau)=\rho_{34}(0) e^{-2\gamma \tau} e^{i \nu_2 \tau}=\frac{1}{25} e^{-2\gamma \tau} e^{i \nu_2 \tau},
\end{equation}
whence the corresponding contributions to the intensity correlation function are $[6/(625 p_5)] e^{-\gamma \tau} \cos(\nu_1 \tau)$, and $[4/(625 p_5)] e^{-2\gamma \tau} \cos(\nu_2 \tau)$,
respectively. The frequencies of the two quantum beats are given by
\begin{subequations}
 \begin{align}
\nu_1&=2g  + \delta_2 - \delta_1, \\
\nu_2&=2\sqrt{2}g  + \delta_4 - \delta_3,
 \end{align}
\end{subequations}
where the frequency shifts $\delta_1, \delta_2, \delta_3, \delta_4$ are calculated in the Appendix. Both these quantum beats have a partial visibility since the coefficients in the superposition states~\eqref{eq:superst} are unequal. Before we continue our discussion on the second-order coherence of the forwards-scattered photons in Sec.~\ref{sec:2ndordcoh}, let us now see what our minimal model can tell us for the phase-space distribution of the intracavity field.  
\begin{figure*}
\centering
\includegraphics[width=0.9\textwidth]{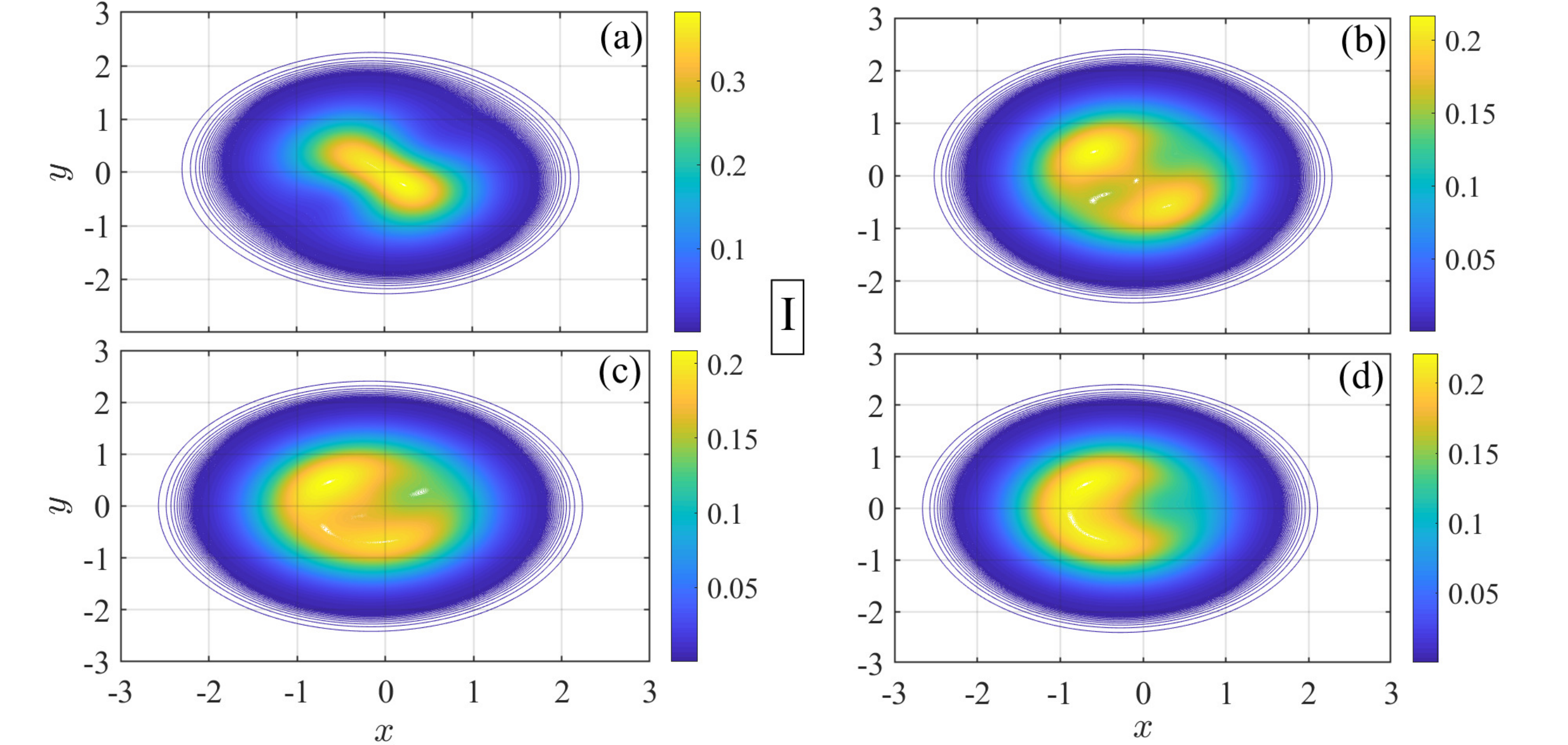}
\includegraphics[width=0.9\textwidth]{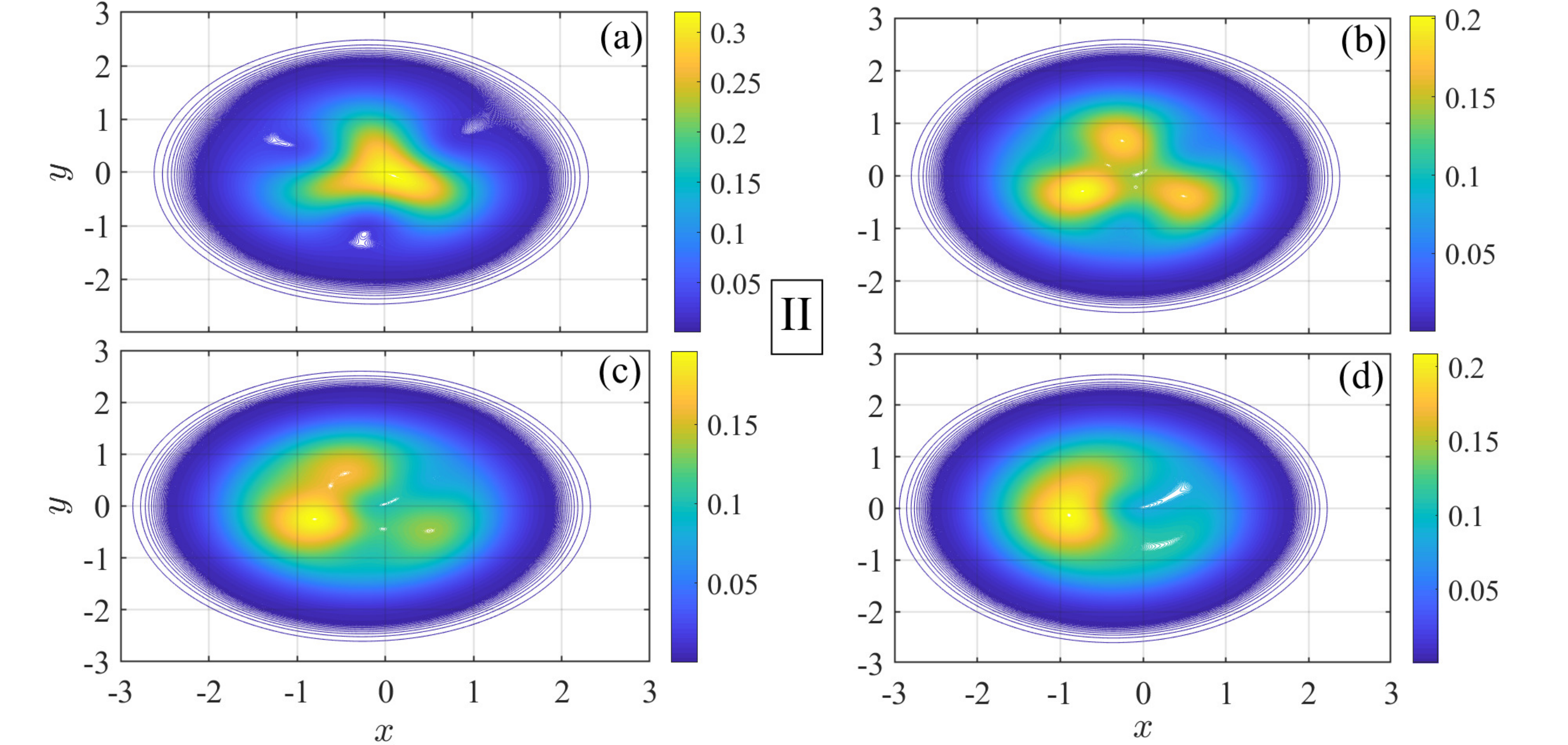}
\caption{{\it Visualizing the breakdown of quantum multimodality in phase space.} Wigner function (contour plots) of the steady-state intracavity field (in the frame rotating with the drive), $W_{\rm ss}(x+iy)$, calculated from the numerical solution of the ME~\eqref{eq:ME1}, depicted in {\bf Panel I} for the {\bf two-photon transition} with $\Delta \omega_d /g=-1/\sqrt{2}-\sqrt{2}(\varepsilon_d/g)^2$, and in {\bf Panel II} for the {\bf three-photon transition} with $\Delta \omega_d/g=-1/\sqrt{3}-(\sqrt{3}/2)(\varepsilon_d/g)^2 + \mathcal{O}[(\varepsilon_d/g)^4]$. In both panels we take $g/\gamma=500$ and $\gamma/(2\kappa)=1$. Frames {\bf (a)}--{\bf (d)} in Panel I correspond to $\varepsilon_d/g=0.02, 0.05, 0.07$ and $0.12$, respectively, while frames {\bf (a)}--{\bf (d)} in Panel II correspond to $\varepsilon_d/g=0.05, 0.07, 0.09$ and $0.125$, respectively. The numerical solution of the ME diagonalizes the Liouvillian $\mathcal{L}$ using an exponential series method, for a Hilbert space of $35$ Fock states in Matlab's {\it Quantum Optics Toolbox}~\cite{Tan_1999}.}
\label{fig:Wss2pb3pb}
\end{figure*}
\begin{figure*}
\centering
\includegraphics[width=0.84\textwidth]{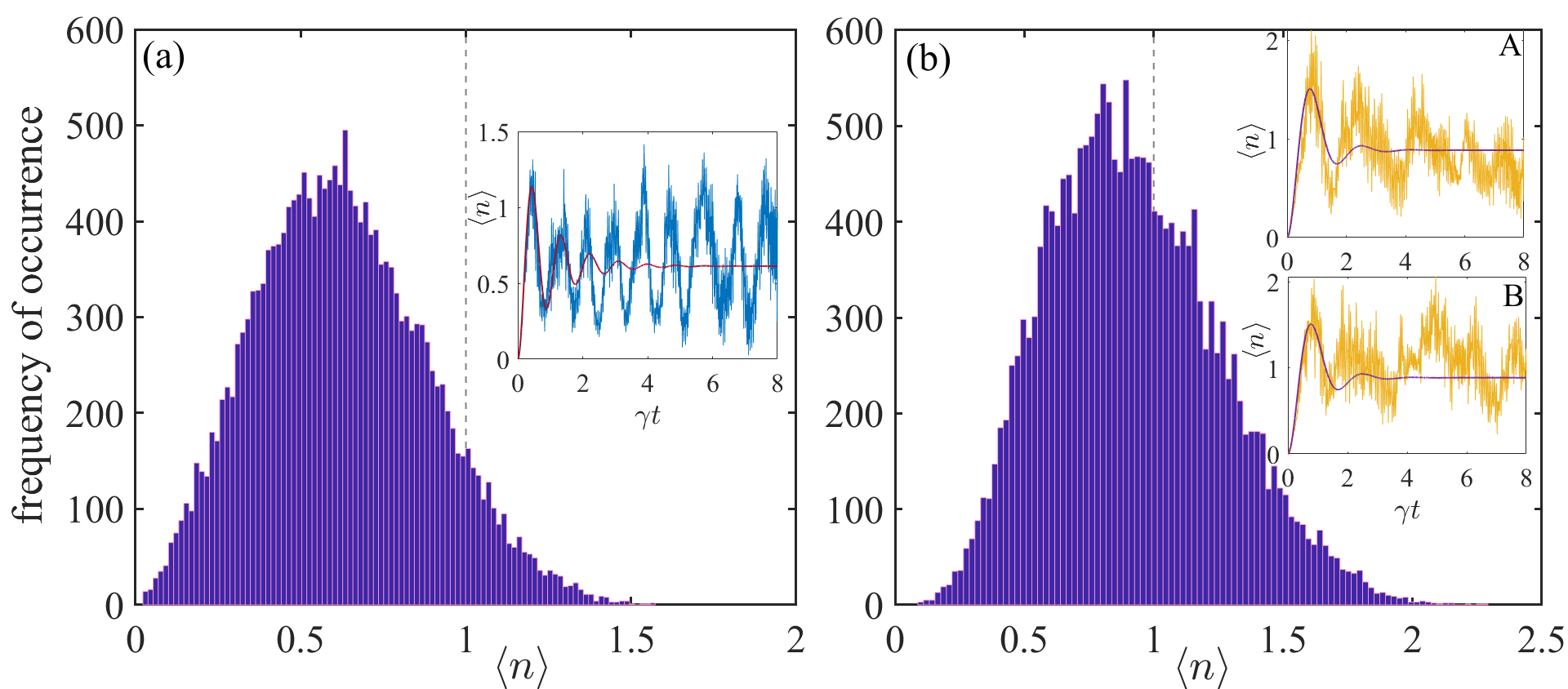}
\caption{{\it Sample quantum trajectories at bi and tri-modality.} Histograms of the time-varying average photon number for the operating conditions used to generate the top-right phase-space profiles in Fig.~\ref{fig:Wss2pb3pb}; the frames (b) of Panels I and II correspond to {\bf (a)} and {\bf (b)}, respectively. The unraveling of the ME~\eqref{eq:ME1} is performed via a quantum state diffusion equation, which is numerically solved by means of a Cash-Karp Runge-Kutta algorithm with adaptive stepsize~\cite{Gisin1992, Schack1997}. A total of $1.92\times 10^4$ data points sampling the time-evolving average $\braket{n(t)}$ in the course of a single trajectory (spanning a time interval $[t_s, t_f]$ with $\gamma t_s=8$ and $\gamma t_f=200$) are split into $100$ bins. The insets depict $\braket{n(t)}=\braket{a^{\dagger}a (t)}$ in the initial stage of the evolution, with initial condition $\braket{n(t=0)}=0$, while the solid lines depict the average photon number obtained from the perturbative treatment for reference -- equivalent to an ensemble average over single trajectories -- in the time interval $[0, t_s)$. In frame (b), the two trajectories A and B correspond to different seeds to the random number generator, and the histogram depicted is extracted from the dataset for A. The dashed gray line marks the value $\braket{n}=1$ in both histograms.}
\label{fig:hist}
\end{figure*}

\section{Steady-state and transient distributions of the cavity field}
\label{sec:WignSST}

The Wigner distribution $W(\alpha, \alpha^{*})$ (in the complex variable $\alpha=x + iy$ and its conjugate, $\alpha^{*}=x-iy$) of an electromagnetic-field state $\rho$ is the Fourier transform of the symmetrically-ordered characteristic function~\cite{CarmichaelQO1}
\begin{equation}
\chi_{{\textstyle\mathstrut}S}(z,z^{*}) \equiv {\rm tr} \left( \rho e^{iz^{*}a^{\dagger}+iza} \right), 
\end{equation}
and is defined as
\begin{equation}
\begin{aligned}
&W(\alpha, \alpha^{*}) \equiv \frac{1}{\pi^2} \int d^2 z\, \chi_{{\textstyle\mathstrut}S}(z,z^{*})\, e^{-iz^{*}\alpha^{*}}\, e^{-iz\alpha}\\
&= \frac{1}{\pi^2} \int_{-\infty}^{\infty}\, d\mu \int_{-\infty}^{\infty} d\nu \, \chi_{{\textstyle\mathstrut}S} (\mu + i\nu, \mu - i\nu)\, e^{-2i(\mu x -\nu y)}.
\end{aligned}
\end{equation}
The reduced density matrix for the cavity field, is defined as $\rho_c \equiv \braket{+|\rho|+} + \braket{-|\rho|-}$. When the steady state is reached it can be recast in the (truncated) Fock-state basis $\ket{n}$ ($n=0, 1, 2, 3$) as [note that the matrix elements $\rho_{ii} \equiv p_i$ ($i=0,1,\ldots,5$) are taken with respect to the dressed-state basis $\ket{\xi_0}-\ket{\xi_5}$]
\begin{equation}
\begin{aligned}
 \rho_{c, {\rm ss}}&=\left[p_0 + \frac{1}{2}(p_1  + p_2)\right]|0 \rangle \langle 0|\\
 &+ \frac{1}{2}(p_1 + p_2 + p_3 + p_4) |1 \rangle \langle 1| +  \frac{1}{2}(p_3 + p_4 + p_5) |2 \rangle \langle 2|\\
 &+ \frac{1}{2}p_5 |2 \rangle \langle 2| + \frac{1}{\sqrt{2}} \left(\rho_{05} |0 \rangle \langle 3| + \rho_{05}^{*} |3 \rangle \langle 0|\right).
\end{aligned}
 \end{equation}
Expressing the coefficients in terms of the steady-state probability $p_5$, we then obtain [compare also with Eq. (14) of~\cite{Wigner2PB}]
\begin{equation}\label{eq:WssSA}
\begin{aligned}
 &W_{\rm ss}(\alpha, \alpha^{*})=\frac{2}{\pi}e^{-2|\alpha|^2}\Bigg\{(1-4 p_5) - \frac{9}{4}p_5 L_1(4|\alpha|^2) \\&+ \frac{5}{4}p_5 L_2(4|\alpha|^2) -\frac{1}{2}p_5 L_3(4|\alpha|^2)\\
 &+i\frac{2}{\sqrt{3}}\sqrt{p_5(4-26p_5)} [\alpha^3-(\alpha^{*})^3]\Bigg\},
 \end{aligned}
\end{equation}
where $L_n$ is the Laguerre polynomial of degree $n$. The last term in the above expression breaks the azimuthal symmetry, which is asymptotically restored as $p_{5} \to 2/13$ (its maximum allowed value). The development of steady-state trimodality is evidenced in Fig.~\ref{fig:Wssanalytical}, based on the analytical expression of Eq.~\eqref{eq:WssSA}; in particular, the term $\propto i[\alpha^3-(\alpha^{*})^3]e^{-2|\alpha|^2}$ dictates the pattern that breaks the azimuthal symmetry of an initial vacuum-state distribution (a Gaussian). This development should be paralleled with Fig. 3 of~\cite{Kerrexact} for the three-photon resonance operation of the driven Kerr oscillator. We note that the three peaks are located outside the unit circle in frames (c, d) of that figure, as does the ring in (e). On the other hand, the intracavity photon number corresponding to all four distributions depicted in Fig.~\ref{fig:Wssanalytical}, equal to $(25/4) p_5$, always remains less than unity (see Fig.~\ref{fig:sat3ph} in the Appendix for a comparison with numerical results). 

For the transient distribution ($\tau >0$) where $\rho_{c}(\tau)\equiv \braket{+|\rho (\tau)|+} + \braket{-|\rho(\tau)|-}$ evolves from the conditional state $\rho_{\rm cond}$, the time-varying coefficients of the various Fock-state contributions, both diagonal and off-diagonal, are  (likewise, the matrix elements $\rho_{ij}$ are taken with respect to the dressed-state basis $\ket{\xi_0}-\ket{\xi_5}$)
\begin{subequations}
 \begin{align}
c_0(\tau)&=\rho_{00}(\tau) + \frac{1}{2}[\rho_{11}(\tau)+ \rho_{22}(\tau)] - \frac{1}{2}[\rho_{12}(\tau)+ \rho_{21}(\tau)], \\
c_1(\tau)&=\frac{1}{2}[\rho_{11}(\tau)+ \rho_{22}(\tau) + \rho_{33}(\tau)+ \rho_{44}(\tau)]\notag\\
&+ \frac{1}{2}[\rho_{12}(\tau)+ \rho_{21}(\tau)-\rho_{34}(\tau)- \rho_{43}(\tau)], \\
c_2(\tau)&=\frac{1}{2}[\rho_{33}(\tau)+ \rho_{44}(\tau) + \rho_{55}(\tau)]+ \frac{1}{2}[\rho_{34}(\tau)+\rho_{43}(\tau)], \\
c_3(\tau)&=\frac{1}{2} \rho_{55}(\tau), \\
c_4(\tau)&=\frac{1}{\sqrt{2}}{\rm Im}[\rho_{05}(\tau)],
 \end{align}
\end{subequations}
where the matrix elements corresponding to the two quantum beats are ($\tau \geq 0$)
\begin{subequations}
 \begin{align}
  \rho_{12}(\tau)&=\rho_{21}^{*}(\tau)=\frac{3}{50}\,e^{-\gamma \tau} e^{i \nu_1 \tau}, \\
  \rho_{34}(\tau)&=\rho_{43}^{*}(\tau)=\frac{1}{25}\,e^{-2\gamma \tau} e^{i \nu_2 \tau}, 
 \end{align}
\end{subequations}
while $\rho_{00}(\tau), \rho_{11}(\tau)+\rho_{22}(\tau), \rho_{33}(\tau)+\rho_{44}(\tau), \rho_{55}(\tau)$ are obtained by numerically solving the system of Eqs.~\eqref{eq:syseq2}. Finally, for $W \equiv W(x+iy;\tau)$, we find
\begin{equation}\label{eq:Wtau}
\begin{aligned}
 &W=\frac{2}{\pi}e^{-2|\alpha|^2}\Big\{c_0(\tau) - c_1(\tau) L_1(4|\alpha|^2)+ c_2(\tau) L_2(4|\alpha|^2)\\& - c_3(\tau) L_3(4|\alpha|^2)+i\,c_4(\tau)(8/\sqrt{6})[\alpha^3-(\alpha^{*})^3]\Big\}.
 \end{aligned}
\end{equation}
 
We are now in position to address our pivotal question from the point of view set by the quantum-classical correspondence: How exactly do quantum fluctuations enter the description of a multiphoton resonance developing along the rungs of the JC ladder? The initial part of our discussion is devoted to the numerical solution of the ME~\eqref{eq:ME1} for the steady state of the intracavity field and its phase-space distributions in the Wigner representation, $W_{\rm ss}(x+iy)$. The first conclusion to be drawn from the profiles depicted in Fig.~\ref{fig:Wss2pb3pb} is that most of the variation associated with the saturation of the multiphoton transition occurs within the unit circle, in line with the analytical {\it quasi}distribution functions of Fig.~\ref{fig:Wssanalytical}; this is a region where the intracavity excitation falls below one photon. The second thing to note is that there is a particular symmetry-breaking mechanism setting in with stronger driving, unique to each multiphoton transition. For the case of the two-photon transition, the initial Gaussian distribution corresponding to the vacuum state is deformed along the line $y=-x$, while for the three-photon transition the probability is distributed along three lines meeting at the origin of the co-ordinates. Equation~\eqref{eq:WssSA} suggests that those lines form an angle of $2\pi/3$ between each other. However, the numerical solution of ME~\eqref{eq:ME1} produces a deviation from that picture, owing to the limited applicability of the secular approximation for higher-order resonances. Subsequently, two and three peaks, respectively, are being formed signaling the presence of bi and trimodality. The development of multimodality is seen from the transition between frames (a) and (b) in the two panels. For a further increase of the drive strength, yet still for $\braket{a^{\dagger}a}_{\rm ss} \leq 1$, the multimodal distributions are erased in favor of a single squeezed state spanning the second and third quadrants in phase space. 

The two-level character of a multiphoton resonance and the mapping to resonance fluorescence when driving the lower vacuum Rabi resonance~\cite{Tian1992} suggest that the photodetection and the updating of post-detection states cannot be built on top of {\it a priori} paths in a classical sense~\cite{OpenSystems2013}. A quantum-trajectory unraveling of the ME~\eqref{eq:ME1} where multimodality is prominent shows that the average photon number at a multiphoton resonance follows rather closely a single-peaked distribution, with its maximum located below the level $\braket{n}=1$. The average values of the two distributions depicted in Fig.~\eqref{fig:hist} are $\overline{\braket{n}}=0.63, 0.92$ for the two and three-photon resonances depicted in frames (a) and (b), respectively [the latter value is that of the left peak in Fig.~\ref{fig:introfig}(c)] independent of the initial state. These results are in agreement with the values of $\braket{a^{\dagger}a}_{\rm ss}$ computed from the steady-state solution to the ME~\eqref{eq:ME1} via diagonalization of the Liouvillian, as expected. More importantly, there is a qualitative difference between the two and the three-photon resonance operation, as revealed by the two insets: the Rabi oscillation at the frequency $2 \Omega \sim \gamma$ noticeable in the first is visibly distorted in the second in terms of its frequency and amplitude, even in the very initial stage of the evolution; despite the fact that the average photon number is higher, the effective three-photon Rabi frequency is lower whence the timescale of multistable switching sets in earlier. Here the quantum beats are absent in the ensemble average  -- a consequence of dephasing in high-frequency oscillations across individual realizations -- since the initial condition (ground state) sets all the corresponding matrix elements to zero. For the quantum beats to feature in the dynamics predicted by the master equation, an appropriate conditional state must be determined by the initial condition; this is indeed the case when we address second-order coherence in Sec~\ref{sec:2ndordcoh}.  

\begin{figure*}
\centering
\includegraphics[width=0.9\textwidth]{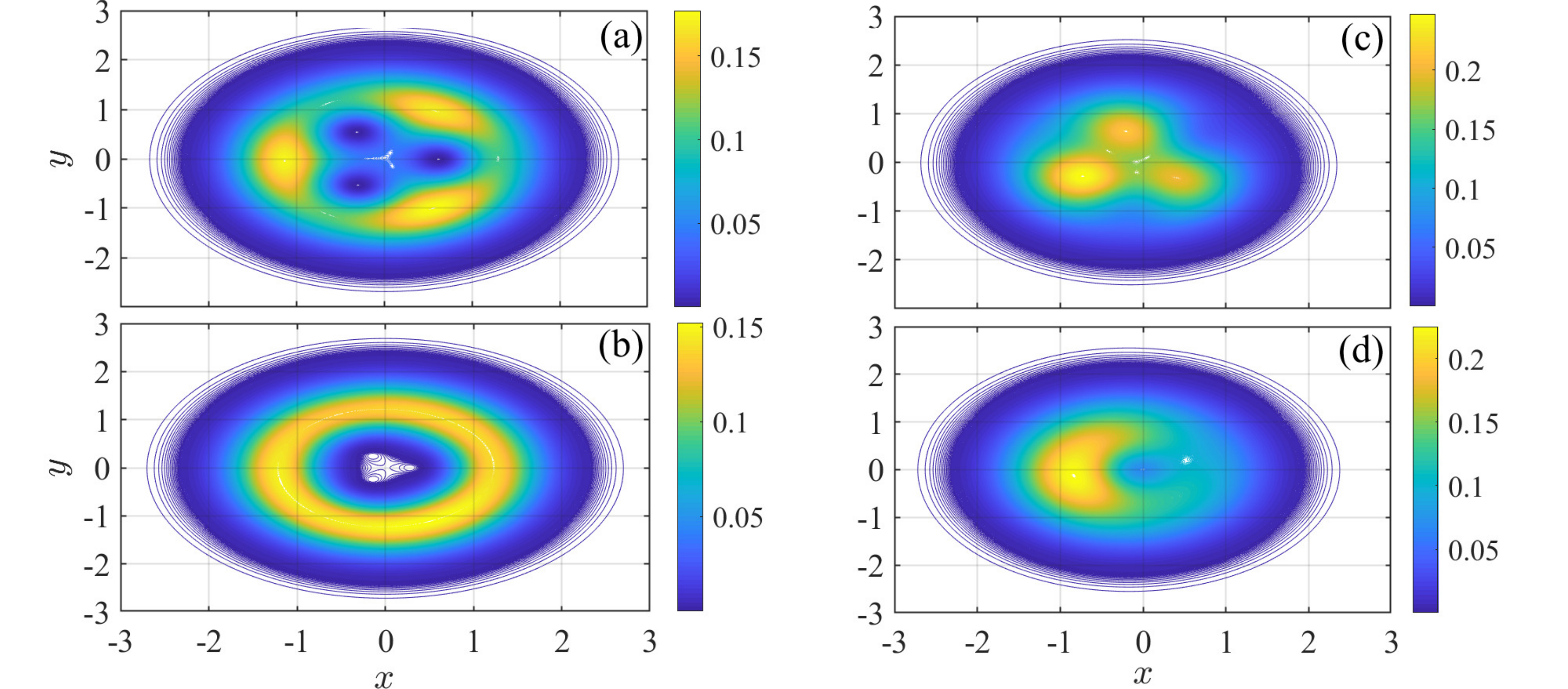}
\includegraphics[width=0.86\textwidth]{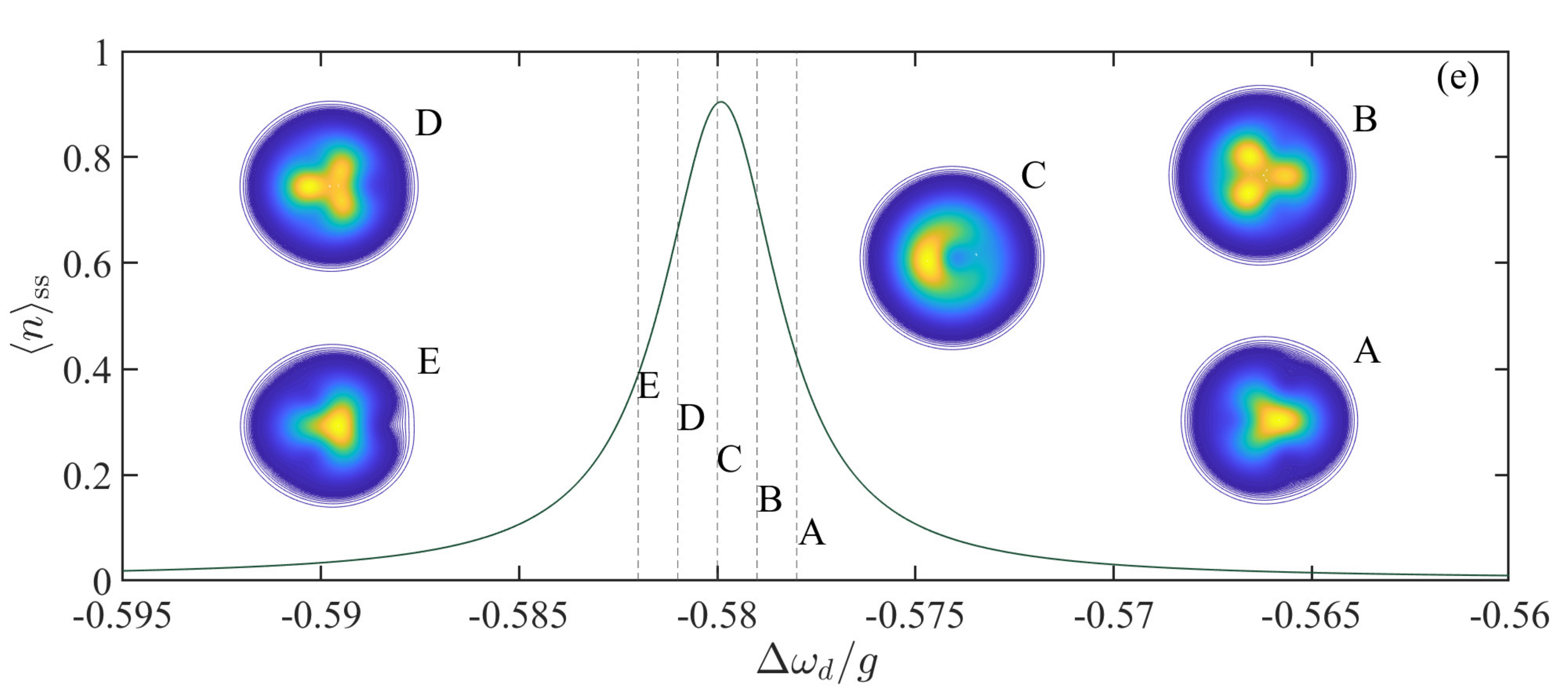}
\caption{{\it Collapse of multimodality in the ``thermodynamic limit'' at zero system size.} Steady-state Wigner distributions (contour plots) of the intracavity field in the Kerr model, extracted from the numerical solution to Eq.~\eqref{eq:KerrME} in {\bf (a)},  {\bf (b)}, and of the JC model, extracted from the numerical solution to Eq.~\eqref{eq:ME1} in {\bf (c)}, {\bf (d)}. In the first case, $\Delta\omega_{d, {\rm K}}/\chi=2$, $|\varepsilon_{d, {\rm K}}|/\chi=0.04$ and $\kappa_{{\rm K}}/\chi=10^{-5}, 10^{-6}$ in (a), (b), respectively; in the second case, $\varepsilon_d/g=0.054$, $\kappa/g=10^{-3}, 10^{-4}$ in (c), (d), respectively, while the cavity detuning is $\Delta\omega_d/g=-1/\sqrt{3}-(\sqrt{3}/2)(\varepsilon_d/g)^2 + \mathcal{O}[(\varepsilon_d/g)^4]$ -- still in very good agreement with the position of the peaks numerically determined by plotting $\braket{n}_{\rm ss}$ vs. $\Delta\omega_d/g$ -- and $\gamma/(2\kappa)=0$. {\bf (e)} The three-photon JC resonance peak against the dimensionless detuning, $\Delta \omega_d/g$, for $\varepsilon_d/g=0.054$, $\kappa/g=10^{-4}$ and $\gamma/(2\kappa)=0$. Schematic profiles of the Wigner function (contour plots) are plotted for the detunings A, B, C, D, E, as indicated in the plot. A Hilbert space of $35$ Fock states has been used for the numerical diagonalization of the corresponding Liouvillian super-operators in Quantum Optics Toolbox.}
\label{fig:KerrandJC}
\end{figure*}

Let us here make a brief detour to revisit the exemplary Kerr nonlinearity. As we have already mentioned above, a trimodal phase-space profile merging to a single-peaked distribution can be analytically derived for the driven Kerr oscillator as an exact steady-state result, using a method based on the complex $P$ representation~\cite{Drummond1980}. Figure 3 of~\cite{Kerrexact} concentrates on the small system size of the oscillator where excitations on the level of one quantum determine its nonlinear response in two-photon blockade conditions. This can be generalized to the phase-space profile for $n$-photon blockade, showing $n$ peaks and $n$ dips; in Fig. 8 of~\cite{Miranowicz2013}, the deformation of the peaks is attributed to the interference in phase space. The master equation of the driven and damped Kerr model in a frame rotating with the frequency of the drive is~\cite{Kerrexact, Drummond1980}
\begin{equation}\label{eq:KerrME}
\begin{aligned}
 \frac{d\rho}{dt}&=i\Delta\omega_{d, {\rm K}}[a^{\dagger}a, \rho]-i \chi[a^{\dagger 2}a^2,\rho]  + [\varepsilon_{d, {\rm K}}a^{\dagger}-\varepsilon_{d, {\rm K}}^{*}a,\rho]\\
 &+\kappa_{{\rm K}}(2a \rho a^{\dagger}-\rho a^{\dagger}a - a^{\dagger}a\rho),
 \end{aligned}
\end{equation}
where $\Delta\omega_{d, {\rm K}} \equiv \omega_{d, {\rm K}}-\omega_0$ is the cavity detuning and $\chi$ is the nonlinear coupling strength of self-phase modulation, proportional to the third-order susceptibility of the medium inside the cavity. A multiphoton resonance requires an integer value of the detuning $\Delta\omega_{d, {\rm K}}/\chi$, while the system-size parameter is determined by the ratio $\kappa_{{\rm K}}/\chi$. In Figs.~\ref{fig:KerrandJC}(a, b) we plot the steady-state Wigner function of the intracavity field for $\Delta\omega_{d, {\rm K}}/\chi=2$ [as is done in Fig. 3 of~\cite{Kerrexact} -- see also Sec. II of~\cite{Miranowicz2013}] and $\kappa_{{\rm K}}/\chi=10^{-5}, 10^{-6}$, respectively. With a decreasing system-size parameter, the three peaks -- positioned outside the unit circle -- merge to an almost azimuthally symmetric distribution while the dips are displaced towards the origin. The situation is quite different for the zero system-size limit [$\gamma/(2\kappa)=0$, see~\cite{CarmichaelQO2, PhotonBlockade2015}] of the driven and damped JC model, when $g^2/(4\kappa^2) \to \infty$ -- the strong-coupling ``thermodynamic limit'' -- with $\varepsilon_d/g$ remaining constant. In Figs.~\ref{fig:KerrandJC}(c, d) we observe that for a decreasing $\kappa/g$, multimodality collapses in the fashion already depicted in Panel II of Fig.~\ref{fig:Wss2pb3pb}, as the three-photon transition saturates -- its remnants are almost indiscernible. This is in contrast to the analytical prediction of Eq.~\eqref{eq:WssSA}, produced for a nonzero $\gamma=2\kappa (\ll g)$, according to which an azimuthally symmetric distribution asymptotically emerges as $\gamma/\Omega \to 0$\footnote{The ratio $\kappa/g$ becomes now the smallest parameter to determine the perturbative expansion. When that ratio is appreciably smaller than the scaled effective Rabi frequency, $\Omega/g$, deviations from the predictions of the ME~\eqref{eq:ME1} are expected. This instance is illustrated by the two Wigner functions of frames (d) in Fig.~\ref{fig:Wss2pb3pb}, which differ from a profile of the kind depicted in Fig.~\ref{fig:Wssanalytical}(d). Nevertheless, trimodality is still evident in Fig.~\ref{fig:KerrandJC}(c) despite the relative smallness of $\kappa/g$.}. Finally, we note that the appearance of multistability is a dynamical effect~\cite{PhotonBlockade2015}; in Fig.~\ref{fig:KerrandJC}(e), we observe the collapse and revival of trimodality as we track the three-photon resonance peak through varying the drive-cavity detuning.  

After having met one notable difference between the JC and Kerr models, both exhibiting multiphoton blockade, we may ask how does the light-matter coupling further modify the manifestation of coherence as the intermediate states $\ket{\xi_1}-\ket{\xi_4}$ in the manifold enter the dynamics? Apart from the deviation of the peak positions in phase space from the particular structure dictated by the drive term in the Hamiltonian of Eq.~\eqref{eq:EffHam}, the answer will be given by bringing into our discussion the photon correlations, and in particular the second-order correlation function of the transmitted light. 

\begin{figure*}
\centering
\includegraphics[width=0.92\textwidth]{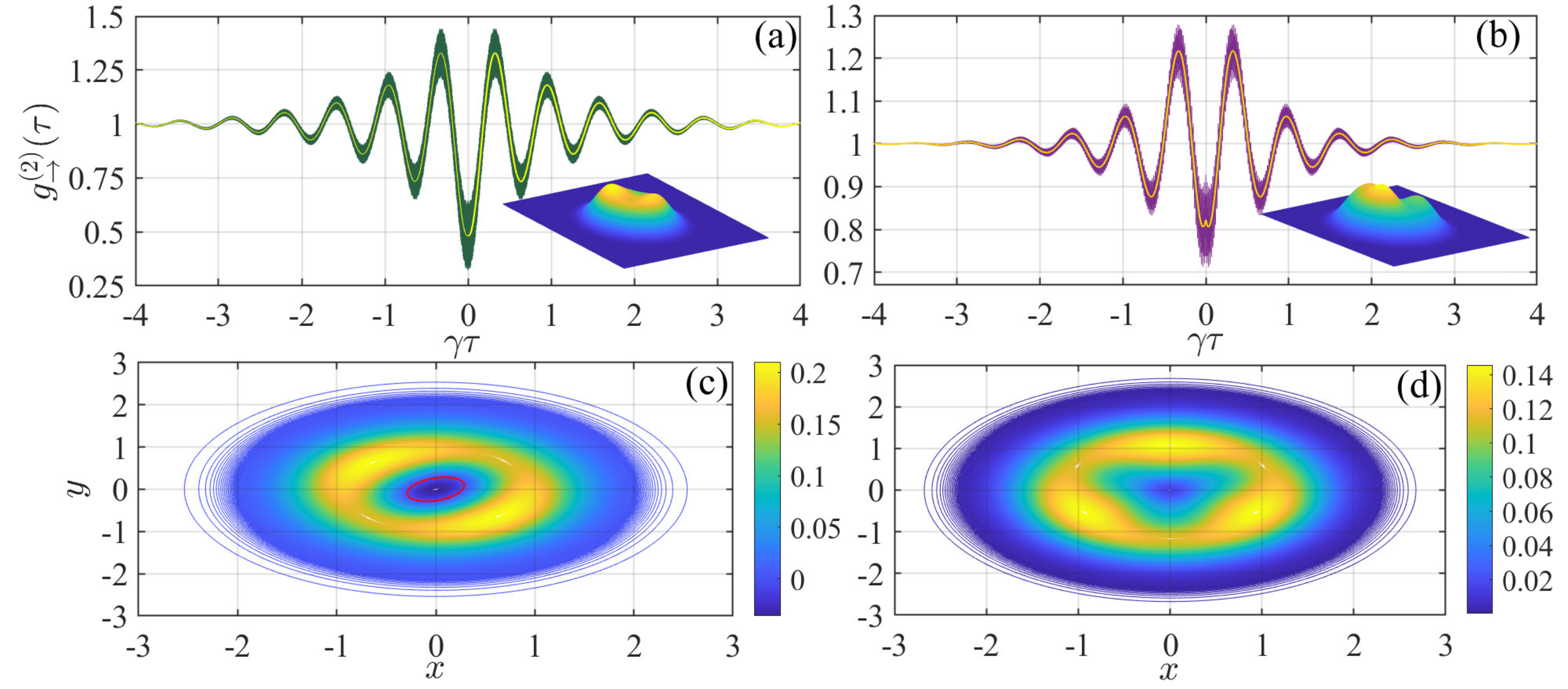}
\includegraphics[width=0.7\textwidth]{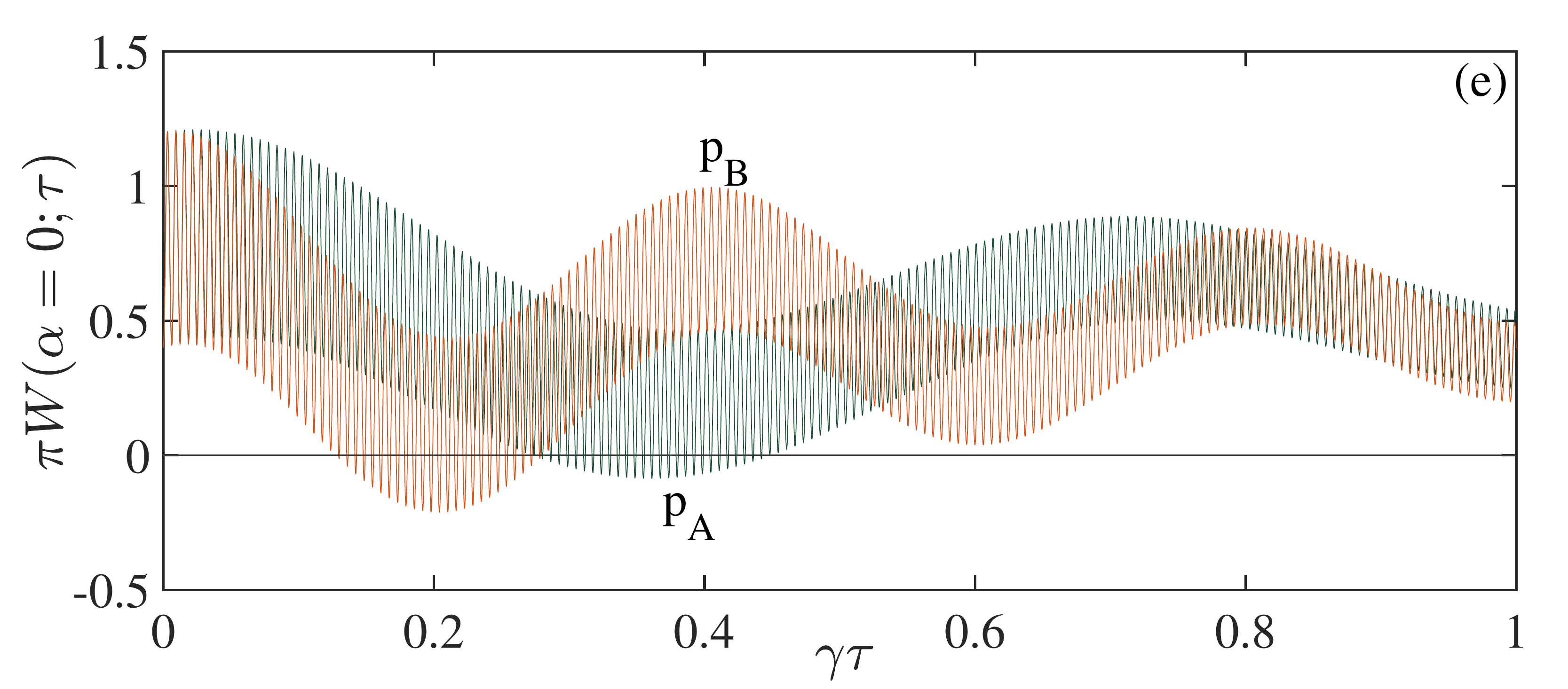}
\includegraphics[width=0.7\textwidth]{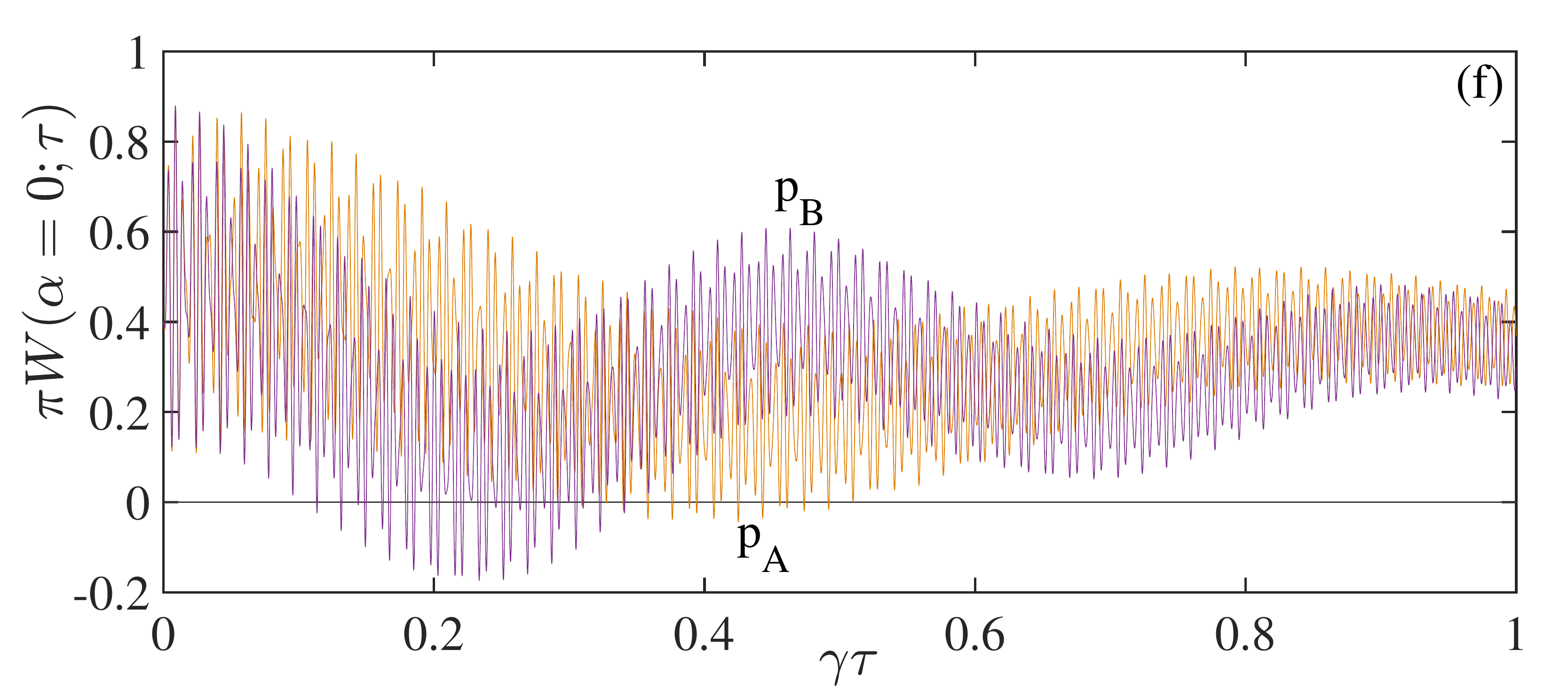}
\caption{{\it Second-order coherence of the transmitted field in multiphoton transitions.} The intensity correlation function $g^{(2)}_{\rightarrow}(\tau)$ for the photons scattered in the forwards direction plotted for the driven {\bf two-photon resonance} in {\bf (a)} is calculated from the analytical expression of Eqs. (44-49) in~\cite{Shamailov2010}, while for the {\bf three-photon resonance} in {\bf (b)} it is numerically computed from our effective six-level model. The solid lines average over the quantum beat(s). The contour plots of the Wigner functions $W(x+iy;\tau_{\rm max})$ in frames {\bf (c)}, {\bf (d)} are determined from the time-varying density matrix $\rho_c (\tau)$ [extracted from the partial trace of $\rho(\tau)$ evolving from $\rho(0)=\rho_{\rm cond}$] at the scaled times $\gamma \tau_{\rm max}$ where the two functions $g^{(2)}_{\rightarrow}(\tau)$ depicted in (a) and (b), respectively, attain their maximum values. They are obtained from Eq.~\eqref{eq:Wtau}. The dashed contour in frame (c) marks out the region inside which the Wigner function attains negative values. The insets in (a), (b) depict surface plots of the corresponding steady-state Wigner functions. For both resonances, we take $\Omega/\gamma=\Omega^{\prime}/\gamma=5$, $g/\gamma=500$ and $\gamma=2\kappa$. {\bf (e)} The time-varying Wigner distribution $W(x=y=0;\tau)$ (multiplied by a factor of $\pi$) of the field state evolving from $\rho(0)=\rho_{\rm cond}$, evaluated at the origin of the phase space and calculated from Eq.~\eqref{eq:W0origin}, for the driven {\bf two-photon resonance}. The two curves correspond to steady-state excitation probabilities $p_A=0.247$ and $p_B=0.249$ of the upper level $\ket{\xi_3}$ (the minimal model comprises the dressed states $\ket{\xi_0}$--$\ket{\xi_3}$), as indicated in the plot. {\bf (f)} Same as in (e), but for the driven {\bf three-photon resonance}. The two curves correspond to steady-state excitation probabilities $p_A=0.1506$ and $p_B=0.1528$ of the upper level $\ket{\xi_5}$ (the minimal model comprises the dressed states $\ket{\xi_0}$--$\ket{\xi_5}$).}
\label{fig:g2andWig}
\end{figure*}

\section{Second-order coherence of the transmitted light}
\label{sec:2ndordcoh}

Our analysis has largely relied on the premise that the two and three-photon resonances are amenable to an effective two-level model~\cite{Tian1992} and the perturbative expansion in powers of $\varepsilon_d/g$, the former more successfully than the latter as the drive detuning to the intermediate transitions gets smaller. For the two-photon transition, where the minimal model comprises the first four levels $\ket{\xi_0}-\ket{\xi_3}$, the analytically determined average photon number in the steady state is $\braket{a^{\dagger}a}_{\rm ss}=(5/2)p_3$, where $p_3=1/[4 + (\gamma/\Omega^{\prime})^2]$ ($\Omega^{\prime}=2\sqrt{2} \varepsilon_d^2/g$) is the steady-state occupation probability of $\ket{\xi_3}$~\cite{Shamailov2010}. For the three-photon transition, the corresponding expression was derived in Sec.~\ref{sec:rateeqs} as $\braket{a^{\dagger}a}_{\rm ss}=(25/4)p_5$, where $p_5=4/[26 + 9(\gamma/\Omega)^2]$ is the steady-state occupation probability of $\ket{\xi_5}$. Both results seem to be reasonable judging from the distributions depicted in Fig.~\ref{fig:Wss2pb3pb} following the numerical solution of the full master equation, since $\braket{a^{\dagger}a}_{\rm ss} \leq 1$ in the distributions where trimodality is present (see also the Appendix). They are also in good agreement with the statistical averages obtained from the trajectory-based histograms we met in Sec.~\ref{sec:WignSST}. Of course, the average photon number only provides a limited account on how quantum fluctuations shape the steady state. Figure~\ref{fig:g2andWig} presents results illustrating the application of the method to the intensity correlation function for the forwards scattered photons, calculated as~\cite{CarmichaelQO2} $g_{\rightarrow}^{(2)}(\tau)= {\rm tr} \{[e^{\mathcal{L}\tau}\rho_{\rm cond}] a^{\dagger}a\}/\braket{a^{\dagger}a}_{\rm ss}$, where $\rho_{\rm cond}$ is the conditional state following the emission of a ``first'' photon from the cavity~\cite{Shamailov2010}, displayed in Eq.~\eqref{eq:rhocondit}. For the two-photon resonance we employ Eqs. (44-49) of~\cite{Shamailov2010} while for the three-photon resonance we numerically solve the system of five coupled equations~\eqref{eq:syseq2}, governing the dynamics of the driven transition and the cascaded decay. 

Let us bring up some additional key results from Sec.~\ref{sec:rateeqs}. When the two-photon transition is excited, the emission of a photon prepares the superposition state $\ket{\xi_3} \to \ket{\psi_{\rm super,\,1}}=\sqrt{2/3}\{[(\sqrt{2}+1)/2]\ket{\xi_1} +  [(\sqrt{2}-1)]/2\ket{\xi_2}]\}$, while for the three-photon transition an additional state is created: $ \ket{\xi_5} \to \ket{\psi_{\rm super,\,2}}=\sqrt{2/5}\{[(\sqrt{3}+\sqrt{2})/2]\ket{\xi_3} + [(\sqrt{3}-\sqrt{2})/2]\ket{\xi_4}\}$. The above states comprise $\rho_{\rm cond}$ alongside a contribution from the vacuum. Evolving in time to determine the intensity correlation of the light emitted from the three-photon cascaded process, they lead to the appearance of two quantum beats, $b_1(\tau)=[6/(625\, p_5)] e^{-\gamma \tau} \cos(\nu_1 \tau)$ and $b_2(\tau)=[4/(625\, p_5)] e^{-2\gamma \tau} \cos(\nu_2 \tau)$, where $\nu_1=2g + \mathcal{O}[(\varepsilon_d/g)^2]$ and $\nu_2=2\sqrt{2}g + \mathcal{O}[(\varepsilon_d/g)^2]$. These terms are responsible for the fast oscillation we observe in frames (a) and (b) of Fig.~\ref{fig:g2andWig} [the beat $\propto e^{-\gamma \tau} \cos(\nu_1 \tau)$ also appears for the two-photon resonance but with a different coefficient~\cite{Shamailov2010}] and provide a direct evidence of the JC spectrum: their frequencies $\nu_1, \nu_2$ are equal to $(\tilde{E}_2-\tilde{E}_1)/\hbar$ and $(\tilde{E}_4-\tilde{E}_3)/\hbar$, respectively, while their coefficients and decay rates show that $b_1$ features more prominently in the correlation than $b_2$; this instance is verified by the Fourier transforms of the intensity correlation function for the fluorescence emitted from the two-level atom, depicted in Fig. 5(b) of~\cite{Lledo2021} [results derived from the solution of the ME ~\eqref{eq:ME1} with $\Delta\omega_d/g=-1/\sqrt{3}$]. While the scope of the secular approximation becomes increasingly limited for multiphoton resonances with $n>3$, we may still state that the beat at frequency $\nu_1$ will outlive higher-order beats, since the coefficients $\Gamma_{ij}$, determining the transition rates between dressed states, scale with the excitation (see Sec. 16.3.3 of~\cite{CarmichaelQO2}). 

On the surface, there is no substantial difference between the intensity correlation functions of Figs. \ref{fig:g2andWig}(a) and \ref{fig:g2andWig}(b). They both exhibit fast oscillations superimposed on top of a varying envelope on the scale $\gamma/\Omega$. A closer look, however, reveals that the range over which the intensity fluctuates is significantly decreased for the three-photon as opposed to the two-photon resonance, which is in turn narrower than the span of $g_{\rightarrow}^{(2)}(\tau)$ for the driven vacuum Rabi resonance (mapped to resonance fluorescence~\cite{CarmichaelQO2, Lang2011}). There is also a difference in the maximum amount of photon antibunching, as predicted by the analytical treatment and verified by the solution of the ME~\eqref{eq:ME1}; while for the two-photon transition we may attain $g_{\rightarrow}^{(2)}(0)=16/25$, for the three-photon resonance we obtain $g_{\rightarrow}^{(2)}(0)=572/625$, which is much closer to unity. For $\Omega=\Omega^{\prime}=5\gamma$, the intensity correlation function in frame (a) has $g_{\rightarrow}^{(2)}(0) \approx 0.65$, compared to $g_{\rightarrow}^{(2)}(0) \approx 0.93$ in frame (b); in fact, for the second case, the exact numerical results from the ME~\eqref{eq:ME1} show weak bunching [$g_{\rightarrow}^{(2)}(0) \approx 1.04$] instead of antibunching. In general, we find no evidence of photon antibunching associated with the collapse of trimodality for the three-photon resonance by applying the ME~\eqref{eq:ME1}, as opposed to the collapse of bimodality in the two-photon resonance where values of $g_{\rightarrow}^{(2)}(0)$ below $1$ are produced for a wide range of $\varepsilon_d/g$. This constitutes further evidence to the increasingly limited scope of the perturbation theory in the secular approximation, when the multiphoton resonances of higher order are reduced to an effective two-state model description.  

Frames (c) and (d) of Fig.~\ref{fig:g2andWig} depict the Wigner function of the field states at which the corresponding intensity correlation functions reach their maximum values. Both distributions show a pronounced dip around the phase-space origin, and for the two-photon resonance this dip extends to negative values -- for the three-photon resonance it doesn't due to the beat interference. Nevertheless, the Wigner function for the three-photon resonance also turns negative in the region around the origin for certain time intervals coordinated with the interference of the two beats. While the interference of the two beats is not resolved on the scale of Fig.~\ref{fig:g2andWig}(b), we find that moving to the immediately adjacent local minimum, occurring $2\pi/\nu_1$ away from the maximum in (a) [or $\sim 4\pi/(\nu_1+\nu_2)$ in (b)], the dip is replaced by a ridge joining the peaks~\cite{Wigner2PB} which are themselves barely noticeable as they form part of an extended ring; there is little resemblance with the steady-state distributions we met in Fig.~\ref{fig:Wss2pb3pb}. 

We close this section out by focusing on the markedly quantum aspect of the above resonances, which is absent from the steady-state distribution. We here observe an imprint of the Fock states, dynamically coordinated with the quantum beat(s), as they participate with varying weights in the transient [note that the term responsible for breaking the symmetry of the phase-space distribution, the last term in Eq.~\eqref{eq:Wtau}, has zero contribution at $x=y=0$]. For simplicity, we consider the two-photon resonance where there is only a single beat. The dynamical evolution is amenable to a four-level minimal model comprising the dressed states $\ket{\xi_0}$--$\ket{\xi_3}$. We can employ our perturbative method to calculate the Wigner function at the origin of the phase space as~\cite{Nogues2000}
\begin{equation}\label{eq:W0origin}
 W(x=y=0; \tau)=\frac{2}{\pi}\sum_{m=0}^{2} (-1)^m \braket{m|\rho_c(\tau)|m},
\end{equation}
where the matrix elements of the reduced density matrix of the cavity field, $\rho_c$, in the four-level model, taken with respect to the Fock states $\ket{m}$, $m=0, 1, 2$, are (omitting the time argument for brevity)
\begin{equation}
\begin{aligned}
 \braket{0|\rho_c|0}&=\rho_{00}-{\rm Re}(\rho_{12}) + \frac{1}{2}(\rho_{11}+\rho_{22}), \\
 \braket{1|\rho_c|1}&= \frac{1}{2}(\rho_{11}+\rho_{22} + \rho_{33}) + {\rm Re}(\rho_{12}), \\
 \braket{2|\rho_c|2}&=\frac{1}{2}\rho_{33}.
 \end{aligned}
\end{equation}
Explicit expressions for the matrix elements $\rho_{ii}$, $i=0, 1, 2, 3$, taken with respect to the dressed-state basis, can be found in Sec. III of~\cite{Wigner2PB}. Figure~\ref{fig:g2andWig}(e) depicts the variation of the transient Wigner distribution, evaluated at the phase-space origin, against the scaled time delay $\gamma \tau$ following detection of one photon. We observe that the saturation of the two-photon resonance (as $p_3 \to p_{3, \, {\rm max}}=0.25$) is accompanied by the appearance of negative values of increasingly larger magnitude, directly coordinated with the quantum beat. Nevertheless, the attainment of negative values around the origin is limited to delays shorter than the coherence time $1/\gamma$. The same is true for the three-photon resonance, as pictured in Fig.~\ref{fig:g2andWig}(f), for which an analogous formula to Eq.~\eqref{eq:W0origin} is used, but now also accounting for the Fock state $\ket{3}$. The interference between the two quantum beats is apparent.   

On the experimental front, we are looking for a connection to be set up between the photon distributions and correlations reported in~\cite{Hamsen2017}, and the phase-space representation of the source field. The scheme devised by Lutterbach and Davidovich in~\cite{Lutterbach1997} is well suited for verifying the alternation of positive and negative values of the Wigner function -- more generally the alternation between a dip and a ridge -- close to the origin of the phase space. In the strong coupling limit and for a given multiphoton transition, this is due to the presence of the quantum beats in the photon correlations superimposed on top of the semiclassical oscillation. The scheme may as well operate across transitions, where the scale of intensity fluctuations shrinks for increasing order. It presents the additional advantage of a direct measurement remaining unaffected by the low detection efficiency, whence it may accurately resolve the depth of the dip displayed in Figs.~\ref{fig:g2andWig}(c-d); the experiment of~\cite{Nogues2000}, implementing the above proposal for a single-photon field, demonstrates the possibility of detecting correlations for low photon averages, namely the regime we are targeting. We may also remark that quantum state diffusion, employed for generating the trajectories depicted in Fig.~\ref{fig:hist}, corresponds to heterodyne detection in particular since one principle used in its construction is the requirement that the added noise terms be independent of phase (see Ch. 18 of~\cite{CarmichaelQO2}). 

\section{Conclusion}

Bringing now the different pieces together, in this article we uncovered the special character of quantum coherence in a multiphoton resonance of the JC model, accompanying the saturation of a driven transition between the ground and an excited dressed state in the spectrum -- dressed by the drive -- for strong dipole coupling conditions. To that end, we developed a perturbative method employing the first few dressed energy levels from the JC ladder, to model the saturation of the driven transition and analytically produce the associated phase-space distribution of the source field. We linked the development of semiclassical Rabi splitting to the collapse of quantum multimodality in the phase-space representation of the intracavity field. The semiclassical oscillations at the effective multiphoton Rabi frequencies are noticeable in the transient of the average photon number. Under operating conditions allowing for the manifestation of trimodality, this number relaxes to a value below or marginally above unity in the steady state, in a background established by the competition of three attractors which are not predicted by mean-field theory -- more accurately, by the coexistence of three excitation pathways followed by intense quantum fluctuations.

Single quantum trajectories, simulating single experimental runs, reveal that the steady-state photon number is established via the ensemble averaging of gradually distorted Rabi oscillations when multimodality sets in. For what is effectively a merge of semiclassical and quantum nonlinearity, the competition of timescales will be determined by the ratio $\Omega/\gamma$. This process, however, is very sensitive to the operating conditions; by further increasing the drive strength multimodality is erased, bringing us closer to a Lorentzian response with a single peak in phase space. Together with the Rabi oscillation in the temporal correlation of forwards scattered photons at the three-photon resonance, we observe an interference between quantum beats whose frequencies directly reveal the excited-state doublets in the JC spectrum. To join the two aspects the manifestation of coherence takes on, we showed that the phase-space distribution of the intracavity field is dynamically associated with the quantum beat, or the interference between two quantum beats, in the transient evolution of the JC model. Finally, despite that steady-state Wigner functions are everywhere positive for all drive strengths, according to the perturbative treatment positive and negative values of the {\it quasi}probability distribution at the origin alternate in a time window set by the decoherence rates. This may occur only during the decoherence time of the transient conditioned on a post-steady-state photodetection.   

\begin{acknowledgments}
I am grateful to C. Lled\'{o} for instructive discussions regarding the application of the perturbative treatment. I acknowledge financial support by the Swedish Research Council (VR) in conjunction with the Knut and Alice Wallenberg foundation (KAW). This work was also supported by the Agencia Estatal de Investigaci\'{o}n (the R\&D project CEX2019-000910-S, funded by MCIN/ AEI/10.13039/501100011033, Plan National FIDEUA PID2019-106901GB-I00, FPI), the Fundaci\'{o} Privada Cellex, Fundaci\'{o} Mir-Puig, and by the Generalitat de Catalunya (AGAUR Grant No. 2017 SGR 1341, CERCA program).
\end{acknowledgments}

\onecolumngrid
\appendix*

\section{Energy shifts and three-photon Rabi frequency following the perturbative treatment}
\label{sec:energyshifts}

In the Appendix, we apply the perturbative method for the determination of the energy shifts arising from the interaction with the drive, which is treated as a small correction to the JC light-matter coupling. Following~\cite{GottfriedQM}, this method was applied by Lled\'{o} to the treatment of the two-photon resonance in~\cite{Lledo2021}; we extend these results here to account for the three-photon resonance. To begin, we decompose the Hamiltonian as $H=H_0 + \varepsilon_d(a+a^{\dagger})$, where 
\begin{equation}
 H_0=-\Delta \omega_d (a^{\dagger}a + \sigma_{+}\sigma_{-}) + g(a\sigma_{+}+a^{\dagger}\sigma_{-}),
\end{equation}
for $\Delta \omega_d=-g/\sqrt{3}$, corresponding to the ``bare'' (not dressed by the drive) three-photon resonance. The Hamiltonian $H_0$ defines the so-called zero-energy subspace of the problem, formed by the states $\ket{\xi_0}$ and $\ket{\xi_5}$ ($E_5=3g/\sqrt{3}-\sqrt{3}g=0=E_0$), and the orthogonal subspace comprising the rest. Treating the term $\varepsilon_d(a+a^{\dagger})$ as a perturbation contributing only small corrections to the dressed energy levels of the JC interaction~\cite{GottfriedQM}, we are led to an expansion for $H_{\rm eff}$ in powers of $\varepsilon_d/g$ which, to the lowest order, is quadratic in the scaled drive strength. Transforming back to the laboratory frame we obtain an expression for $\tilde{H}_{\rm eff}$, which is displayed in Eq.~\eqref{eq:EffHam}. 

We first meet a case where we need to carry the calculation out to higher order than second to obtain the dominant contribution, in contrast to the two-photon resonance. For the effective three-photon drive, following the standard procedure of third-order perturbation theory and summing the contributions from the four de-excitation pathways depicted in Fig.~\ref{fig:6levels}, we obtain:
\begin{align}\label{eq:Omega3p}
\Omega=\braket{\xi_5|H_{\rm eff}|\xi_0} &= \varepsilon_d^3 \sum\limits_{k,l\neq (5,0)} \frac{1}{E_k}\frac{1}{E_l} \braket{\xi_5|(a^\dag + a)|k} \braket{k|(a^\dag + a)|l}\braket{l|(a^\dag + a)|\xi_0}\notag\\
 &=\varepsilon_d^3 \frac{1}{E_4 E_2}\braket{\xi_5|(a^\dag + a)|\xi_4} \braket{\xi_4|(a^\dag + a)|\xi_2}\braket{\xi_2|(a^\dag + a)|\xi_0}\notag\\
 & + \varepsilon_d^3 \frac{1}{E_4 E_1}\braket{\xi_5|(a^\dag + a)|\xi_4} \braket{\xi_4|(a^\dag + a)|\xi_1}\braket{\xi_1|(a^\dag + a)|\xi_0}\notag\\
 &+ \varepsilon_d^3 \frac{1}{E_3 E_1}\braket{\xi_5|(a^\dag + a)|\xi_3} \braket{\xi_3|(a^\dag + a)|\xi_1}\braket{\xi_1|(a^\dag + a)|\xi_0}\notag\\
 & + \varepsilon_d^3 \frac{1}{E_3 E_2}\braket{\xi_5|(a^\dag + a)|\xi_3} \braket{\xi_3|(a^\dag + a)|\xi_2}\braket{\xi_2|(a^\dag + a)|\xi_0},
\end{align}
where $E_1=g/\sqrt{3}-g$, $E_2=g/\sqrt{3}+g$, $E_3=2g/\sqrt{3}-\sqrt{2}g$, $E_4=2g/\sqrt{3}+\sqrt{2}g$. Summing the various contributions yields
\begin{equation}
 \Omega=\frac{3}{4\sqrt{2}} \left[\frac{(\sqrt{3}-\sqrt{2})(1 + \sqrt{2})}{(2+\sqrt{6})(1+\sqrt{3})} + \frac{(\sqrt{3}-\sqrt{2})(\sqrt{2}-1)}{(2+\sqrt{6})(1-\sqrt{3})} + \frac{(\sqrt{3}+\sqrt{2})(\sqrt{2}+1)}{(2-\sqrt{6})(1-\sqrt{3})} + \frac{(\sqrt{3}+\sqrt{2})(\sqrt{2}-1)}{(2-\sqrt{6})(1+\sqrt{3})} \right]\frac{\varepsilon_d^3}{g^2}\approx 11.69 \frac{\varepsilon_d^3}{g^2}.
\end{equation}
We now illustrate the application of the method to the dressing of the JC levels. Here, the dominant-order contributions scale as $\varepsilon_d^2/g$. For the ground state, we find
\begin{equation}
\begin{aligned}
 \delta_0&=\varepsilon_d^2 \sum_{k \neq 0} \frac{1}{-E_k} \braket{\xi_0|(a^{\dagger}+a)|k} \braket{k|(a^{\dagger}+a)|\xi_0}\\
 &=\varepsilon_d^2 \left(-\frac{1}{E_1} \left|\braket{\xi_1|(a^{\dagger}+a)|\xi_0}\right|^2 -\frac{1}{E_2} \left|\braket{\xi_2|(a^{\dagger}+a)|\xi_0}\right|^2 \right)\\
 &=\frac{\varepsilon_d^2 \sqrt{3}}{2g}\left(\frac{1}{\sqrt{3}-1} - \frac{1}{\sqrt{3}+1}\right)\\
 &=\frac{\sqrt{3}}{2}\frac{\varepsilon_d^2}{g}.
 \end{aligned}
\end{equation}
Likewise, for $\ket{\xi_1}$ we obtain
\begin{equation}
\begin{aligned}
 \delta_1&=\varepsilon_d^2 \sum_{k \neq 1} \frac{1}{E_1-E_k} \braket{\xi_1|(a^{\dagger}+a)|k} \braket{k|(a^{\dagger}+a)|\xi_1}\\
 &=\frac{\varepsilon_d^2}{g} \left[\frac{1}{2}\frac{\sqrt{3}}{1-\sqrt{3}} + \left(\frac{\sqrt{2}+1}{2}\right)^2\frac{\sqrt{3}}{\sqrt{6}-\sqrt{3}-1} - \left(\frac{\sqrt{2}-1}{2}\right)^2\frac{\sqrt{3}}{\sqrt{3}+1+\sqrt{6}} \right],
 \end{aligned}
\end{equation}
while for $\ket{\xi_2}$, 
\begin{equation}
\begin{aligned}
 \delta_2&=\varepsilon_d^2 \sum_{k \neq 2} \frac{1}{E_2-E_k} \braket{\xi_2|(a^{\dagger}+a)|k} \braket{k|(a^{\dagger}+a)|\xi_2}\\
 &=\frac{\varepsilon_d^2}{g} \left[\frac{1}{2}\frac{\sqrt{3}}{1+\sqrt{3}} + \left(\frac{\sqrt{2}-1}{2}\right)^2\frac{\sqrt{3}}{\sqrt{6}+\sqrt{3}-1} + \left(\frac{\sqrt{2}+1}{2}\right)^2\frac{\sqrt{3}}{\sqrt{3}-1-\sqrt{6}} \right].
 \end{aligned}
\end{equation}
For the second-excited doublet,
\begin{equation}\label{eq:delta3}
\begin{aligned}
 \delta_3&=\varepsilon_d^2 \sum_{k \neq 3} \frac{1}{E_3-E_k} \braket{\xi_3|(a^{\dagger}+a)|k} \braket{k|(a^{\dagger}+a)|\xi_3}\\
 &=\frac{\varepsilon_d^2}{g} \Bigg[\left(\frac{\sqrt{3}+\sqrt{2}}{2}\right)^2\frac{\sqrt{3}}{2-\sqrt{6}} - \left(\frac{\sqrt{3}-\sqrt{2}}{2}\right)^2\frac{\sqrt{3}}{4+\sqrt{6}}\\
 &+ \left(\frac{\sqrt{2}+1}{2}\right)^2\frac{\sqrt{3}}{1+\sqrt{3}-\sqrt{6}} + \left(\frac{\sqrt{2}-1}{2}\right)^2\frac{\sqrt{3}}{1-\sqrt{3}-\sqrt{6}} \Bigg],
 \end{aligned}
\end{equation}
\begin{equation}\label{eq:delta4}
\begin{aligned}
 \delta_4&=\varepsilon_d^2 \sum_{k \neq 4} \frac{1}{E_4-E_k} \braket{\xi_4|(a^{\dagger}+a)|k} \braket{k|(a^{\dagger}+a)|\xi_4}\\
 &=\frac{\varepsilon_d^2}{g} \Bigg[\left(\frac{\sqrt{3}-\sqrt{2}}{2}\right)^2\frac{\sqrt{3}}{2+\sqrt{6}} + \left(\frac{\sqrt{3}+\sqrt{2}}{2}\right)^2\frac{\sqrt{3}}{\sqrt{6}-4}\\
 &+ \left(\frac{\sqrt{2}+1}{2}\right)^2\frac{\sqrt{3}}{1+\sqrt{6}-\sqrt{3}} + \left(\frac{\sqrt{2}-1}{2}\right)^2\frac{\sqrt{3}}{1+\sqrt{3}+\sqrt{6}}\Bigg].
 \end{aligned}
\end{equation}
The second terms in Eqs.~\eqref{eq:delta3} and \eqref{eq:delta4} originate from transitions involving the ``bare'' state $\ket{\xi_6} \equiv (1/\sqrt{2})(\ket{3,-} + \ket{2,+})$, which are assumed to be far from resonance.  Finally,
\begin{equation}
\begin{aligned}
 \delta_5&=\varepsilon_d^2 \sum_{k \neq 5} \frac{1}{E_5-E_k} \braket{\xi_5|(a^{\dagger}+a)|k} \braket{k|(a^{\dagger}+a)|\xi_5}\\
 &=\frac{\varepsilon_d^2}{g} \Bigg[\left(\frac{\sqrt{3}+\sqrt{2}}{2}\right)^2\frac{\sqrt{3}}{\sqrt{6}-2} - \left(\frac{\sqrt{3}-\sqrt{2}}{2}\right)^2\frac{\sqrt{3}}{\sqrt{6}+2}\\
 &+ \left(\frac{2+\sqrt{3}}{2}\right)^2\frac{\sqrt{3}}{2\sqrt{3}-4} - \left(\frac{2-\sqrt{3}}{2}\right)^2\frac{\sqrt{3}}{4+2\sqrt{3}}\Bigg]= -\sqrt{3} \frac{\varepsilon_d^2}{g}.
 \end{aligned}
\end{equation}
\begin{figure}
\begin{center}
\includegraphics[width=0.75\textwidth]{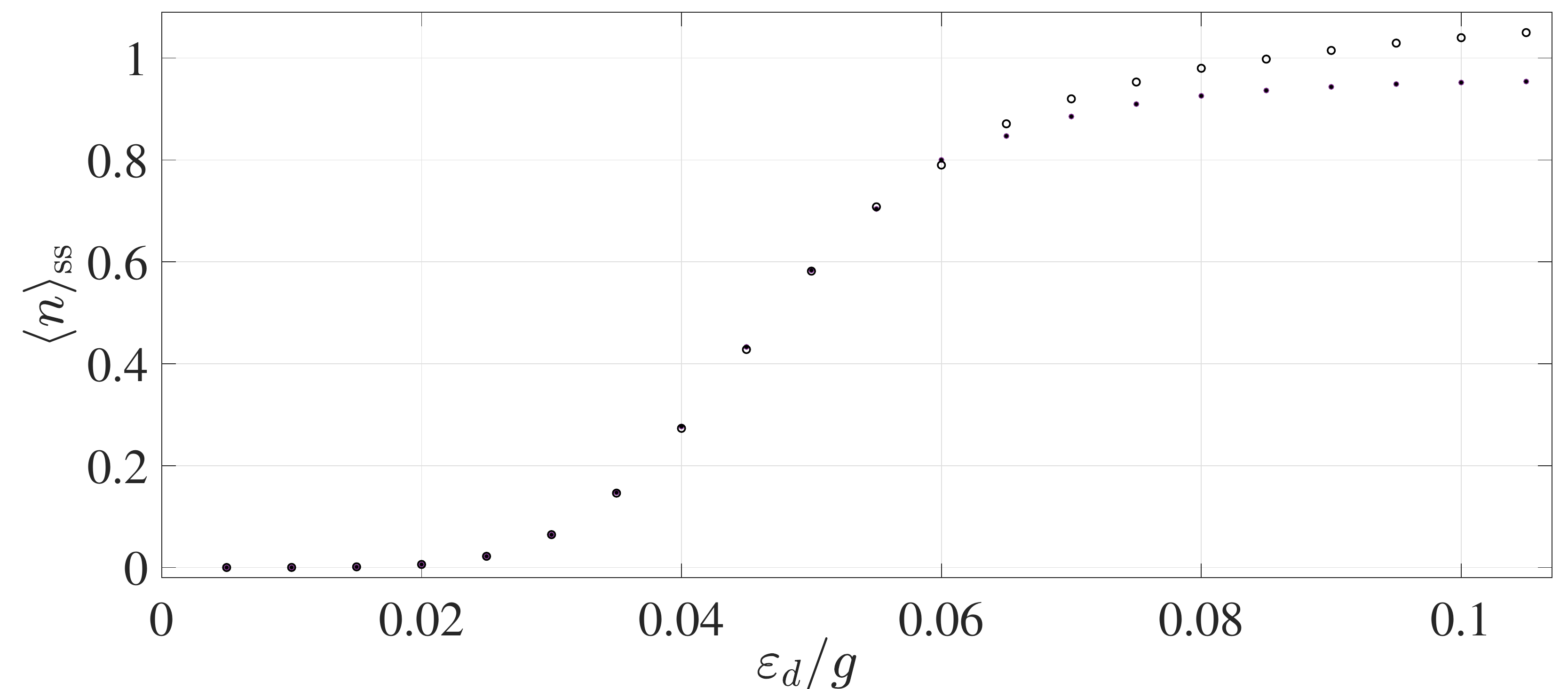}
\end{center}
\caption{{\it Saturation of the three-photon resonance.} Steady-state photon number $\braket{n}_{\rm ss}$ at the three-photon resonance peak, occurring for $\Delta \omega_d/g=-1/\sqrt{3}-(\sqrt{3}/2)(\varepsilon_d/g)^2 + \mathcal{O}[(\varepsilon_d/g)^4]$, as a function of the scaled drive strength $\varepsilon_d/g$. The results obtained from the numerical solution of the ME~\eqref{eq:ME1} (open circles) are plotted against the analytical perturbative expression $\braket{n}_{\rm ss}=25/[26+9(\gamma/\Omega)^2]$, derived from the six-state model (filled circles). The three-photon Rabi frequency $\Omega$ is derived from Eq.~\eqref{eq:Omega3p} as a perturbative shift to the zero-energy subspace in the rotated frame. We take $g/\gamma=500$ and $\gamma=2\kappa$.}
\label{fig:sat3ph}
\end{figure}

To improve even further the agreement with the exact numerical results obtained from the solution of the ME~\eqref{eq:ME1}, the difference $\delta_5 - \delta_0$ in Eq.~\eqref{eq:rescond} could be calculated to higher order, accounting for the shifts $\sim \varepsilon_d^2/g$ in the detuning from intermediate levels. This procedure leaves us with dressed energy-level shifts $\tilde{\delta}_{(0,5)}$ satisfying $
\tilde{\delta}_{(0,5)} -\delta_{(0,5)} \propto [(\delta_{i \neq (0,5)}-\delta_{(0,5)})/g]\,(\varepsilon_d^2/g) \sim \varepsilon_d^4/g^3$, while it generates corrections of order $(\varepsilon_d/g)^5$ to $\Omega/g$. As a general rule, for the perturbative treatment to remain valid, we require that the ratio $(\tilde{\delta}_5 - \tilde{\delta}_0)/\tilde{\Omega} \sim 1$ as well as $\tilde{\Omega}/\gamma \sim 1$ i.e., that the dressed Rabi frequency ($\tilde{\Omega}$), the energy shifts and the dissipation rates are of the same order of magnitude. The driving term (perturbation) together with the form of the ``bare'' JC states require that the effective Rabi frequency of the $n$-photon transition scale as $\varepsilon_d^n/g^{n-1}$, while the level shifts always follow $\varepsilon_d^2/g$ to dominant order. 

Comparison with numerical results shows that the perturbative energy shifts featuring in Eq.~\eqref{eq:rescond} reproduce very accurately the peak positions of the three-photon resonance, even for $\gamma/(2\kappa) \to 0$, while the average photon number is underestimated past a certain value of $\varepsilon_d/g$, since the perturbative result is constrained to remain below unity, bounded by $25/26$. Figure~\ref{fig:sat3ph} depicts the saturation of the three-photon resonance, comparing the numerical and analytical results. In addition, the numerical results show that the zero-delay correlation function $g_{\rightarrow}^{(2)}(0)$ never falls below unity in the drive region used for Fig.~\ref{fig:sat3ph}, in contrast to the analytical prediction of Eq.~\eqref{eq:g20analytical}. According to this prediction, photon antibunching in the forwards direction is expected for $88/625<p_5<2/13$. Finally, we mention that the presence of dissipation induces a coupling between the zero-energy and its orthogonal subspace of dressed JC states; this coupling is neglected on account of the secular approximation.

\twocolumngrid
\bibliography{bibliography}

\end{document}